\pdfoutput=1
\documentclass[12pt,onecolumn,draftclsnofoot,journal]{IEEEtran}
\usepackage[margin=2.54cm, lmargin=2.54cm]{geometry}
\usepackage{amsthm}
\usepackage{amssymb}
\usepackage{latexsym}
\usepackage{amsfonts}
\usepackage{amsbsy}
\usepackage{amsmath,amssymb}
\usepackage{times}
\usepackage{graphicx}
\usepackage{enumerate}
\usepackage[usenames]{color}
\usepackage[dvips]{pstcol}
\usepackage{epstopdf}
\usepackage{subfig}
\usepackage{bm}
\usepackage{booktabs}
\usepackage{cite}
\usepackage{color}
\usepackage{xcolor}
\usepackage{algorithm}
\usepackage{algorithmicx}
\usepackage{algpseudocode}
\usepackage{setspace}
\usepackage{bm}
\usepackage{tabularx}
\usepackage{amstext,epsfig,graphicx}
\usepackage{psfrag}
\usepackage{cite}
\usepackage{footnote}
\usepackage{siunitx}
\usepackage{mathrsfs}
\usepackage{booktabs}
\usepackage{multirow}
\usepackage{stfloats}

\begin{document}

%
\title{Channel Estimation for IRS-aided Multiuser Communications with Reduced Error Propagation}
%
%
%
\vspace{-0.3cm}
\author{Yi Wei, Ming-Min Zhao, Min-Jian Zhao, and Yunlong Cai \vspace{-0.7cm}
\thanks{The authors are with the College of Information Science and Electronic Engineering, Zhejiang University, Hangzhou 310027, China
(email: \{21731133, zmmblack, mjzhao, ylcai\}@zju.edu.cn).} 


%

}
\IEEEpeerreviewmaketitle

\vspace{-1.5em}

\maketitle

\vspace{-2.8em}
\begin{abstract}
\begin{spacing}{1.46}
Intelligent reflecting surface (IRS) has emerged as a promising  paradigm to improve the capacity and reliability of a wireless communication system  by smartly reconfiguring the wireless propagation environment.
To achieve the promising gains of IRS, the acquisition of the channel state information (CSI) is essential, which however is practically difficult since the IRS does not employ any transmit/receive radio frequency (RF) chains in general and it has limited signal processing capability. 
In this paper, we study the uplink channel estimation problem for an IRS-aided multiuser single-input multi-output (SIMO) system.
{ The existing  channel estimation approach for IRS-aided multiuser systems mainly consists of three phases, where  the  direct channels from the base station (BS) to all the users, the reflected channel from the BS to a typical user via the IRS, and the other reflected channels are estimated sequentially based on the estimation results of the previous phases.
However, this approach will lead to a serious error propagation issue, i.e., the channel estimation errors in the first and second phases will deteriorate the estimation performance in the second and third phases.}
To resolve this difficulty, 
we propose a novel two-phase  channel estimation (2PCE)  strategy
which is able to alleviate  the negative effects caused by error propagation and enhance the channel estimation performance with the same amount of channel training overhead as in the existing approach. Specifically, in the first phase, the direct and reflected channels associated with a typical user are estimated simultaneously by varying the reflection patterns at the IRS, such that the estimation errors of the direct channel associated with this typical user will not affect the estimation of the corresponding reflected channel. In the second phase, we estimate the CSI associated with the other users and demonstrate that by properly designing the pilot symbols of the users and the reflection patterns at the IRS, the direct and reflected channels associated with each user can also be estimated simultaneously, which helps to reduce the error propagation.
Moreover,  the asymptotic mean squared error (MSE) of the proposed 2PCE strategy is  analyzed when the least-square (LS) channel estimation method is employed, and we show that the 2PCE strategy can outperform the existing approach.
Finally, extensive simulation results are presented to  validate the effectiveness of our proposed channel estimation strategy.
\end{spacing}
\end{abstract}

\vspace{-0.6cm}
\begin{IEEEkeywords}
Intelligent reflecting surface, channel estimation, multiuser communications, single-input multiple-output (SIMO).
\end{IEEEkeywords}

\vspace{-0.4cm}
\linespread{1.5}
\begin{spacing}{1.46}
\vspace{-0.3cm}\section{Introduction}
To  meet the ever-increasing  demand for high-speed  and low-latency data applications, various advanced technologies,  e.g.,  massive multiple-input  multiple-output (MIMO) communication, millimeter wave (mmWave) communication and
ultra-dense network (UDN), have been proposed and extensively studied to improve the spectral efficiency of wireless systems \cite{Boccardi2014,Shafi2017}.
These technologies have led to fundamental changes in the design of the fifth generation (5G) cellular network and brought significant performance improvement, however, they also incur high energy consumption and hardware cost, due to the large number of active nodes/antennas/radio-frequency (RF) chains used. Recently, intelligent reflecting surface (IRS) has been proposed as a promising new paradigm  to alleviate the above issues and enhance the system performance \cite{Wu2020,Basar2019,Qing2020,Tan2018}.
In general, IRS is a man-made  planar surface  consisting of a large number of passive and low-cost reflecting elements, each of which can be independently adjusted to change the amplitude and/or phase of the  incident signal by an external software controller. As a result, by smartly coordinating  these  reflecting elements, IRS is able to  create a favorable wireless signal propagation environment to  improve
the wireless communication
coverage, throughput, and energy efficiency substantially.
Besides, different from the conventional half-duplex amplify-and-forward (AF) relay, IRS can  achieve high beamforming gains by intelligently adjusting its reflection coefficients at different reflecting elements in a full-duplex manner, with low hardware and energy cost.

As such, IRS has attracted significant attention recently and various works have been conducted in the literature to show the effectiveness of IRS in enhancing the system performance, see e.g., \cite{MingMin2020,Larsson2020,Aliu,RZhang2019}.  However,
to reap the performance gains offered by IRS, the acquisition of channel state information (CSI) is of significant importance since the IRS is generally not equipped with any transmit/receive RF chains and thus not capable of performing complex baseband signal processing tasks.
{This issue becomes more stringent when the numbers of base station (BS) antennas, IRS reflecting elements and users are large, since
the number of channel coefficients that are required to be estimated will become considerably large. }
Although there are several studies in the literature  proposed to design the IRS passive beamforming based on the statistical CSI and employ the two-timescale transmission protocol to reduce the channel training overhead \cite{MMZhao2}, how to estimate the full CSI efficiently is still an essential problem since various  works require full CSI for IRS reflection optimization and obtaining a larger number of channel samples is also an indispensable part for statistical CSI estimation.

Recently, various strategies have been proposed to tackle the channel estimation problem efficiently for IRS-aided wireless communication systems \cite{Mishra2019,9039554,Jensen2019,OFDM2020,Tellambura2019,CYou2019,He2019,Jie2019,9149146,9214532,arXiv1912}. Specifically,
for the single-user system,
the studies of  \cite{Mishra2019},{\cite{9039554}} proposed binary reflection controlled channel estimation schemes, where  the user-BS channel (referred to as the \emph{direct channel}) is estimated first when all the reflecting elements are turned off, then  only one reflecting
element is switched on at each subsequent time slot to estimate the cascaded user-IRS-BS channel (referred to as the \emph{reflected channel}) associated with this reflecting element. In this strategy, as only a small portion of the IRS reflecting elements is switched on at each time, the channel estimation accuracy may be degraded as the IRS's large aperture gain is not exploited. To address this issue, the works \cite{Jensen2019} and \cite{OFDM2020} improved the binary reflection strategy by considering the full reflection of the IRS at all time slots and designing a discrete Fourier transform (DFT) based IRS training reflection matrix  (also known as \emph{reflection pattern}). 
  Further, the work \cite{CYou2019} studied the reflection pattern design problem under the practical constraint that the phase shift at each reflecting element only takes a finite number of discrete values.
  When massive MIMO system is considered, the authors in \cite{He2019} exploited the low-rank structure of the channel matrix and formulated  the reflected channel estimation problem as a combined sparse matrix factorization and matrix completion problem. {As for the IRS-aided mmWave MIMO systems, the works \cite{9149146,9214532} studied the channel estimation problem  by exploiting the channel sparsity.}  On the other hand, for the multiuser system, the work \cite{Jie2019}
 proposed a compressive sensing (CS) based channel estimation method with reduced channel training overhead, based on the assumption that the reflected channels are sparse.
  For more general channel models, the work \cite{arXiv1912} proposed a novel three-phase channel estimation (3PCE) framework to estimate the direct and reflected channels, where the
 strong correlation among the reflected channels associated with different users
 is exploited to reduce the channel training overhead.
Specifically, since each IRS reflecting  element reflects the signals from different users to
the BS via the same IRS-BS channel, the reflected channel of an arbitrary user is a scaled version of that of the other users and only the scaling factors (scalars), rather than the whole channels (vectors), are needed to be estimated.
 Under this framework, only $K+N+\max(K-1,\lceil\frac{(K-1) N}{M}\rceil)$ pilot symbols are required to estimate the full CSI which consists of $KM+KMN$ channel coefficients, instead of $K+KN$ pilot symbols if this correlation is not considered, where $K$, $N$ and $M$ denote the numbers of users, IRS reflecting elements, and BS antennas, respectively.
 However, these exist a serious error propagation issue in this 3PCE framework, i.e., the channel estimation errors in the first and second phases will deteriorate the estimation performance in the second and third phases.

 Motivated by the above, we consider an IRS-aided multiuser uplink SIMO communication system in this paper where multiple single-antenna users communicate with a multi-antenna BS with the assist of an IRS, and propose a new channel estimation  strategy to alleviate the error propagation issue in the existing 3PCE strategy \cite{arXiv1912}.  Our main idea is to reduce the channel estimation phases by estimating the direct and reflected channels associated with each user simultaneously, such that the error propagation between different phases can be well-controlled. Specifically, our proposed strategy { which is referred to as the two-phase channel estimation (2PCE) strategy } mainly consists of two estimation phases, where the first phase aims to estimate the direct and reflected channels associated with a typical user and the second phase is dedicated to estimate the other channels based on the knowledge of this typical user's CSI. We first show that the proposed 2PCE strategy is able to perfectly  estimate all the channel coefficients with the theoretically minimal pilot sequence length, under the ideal case without receive noise at the BS. Then, for the practical case with receive noise at the BS,
the least-square (LS) channel estimation solutions are derived in both phases.
  The  asymptotic MSE of the proposed 2PCE strategy is analyzed when the number of transmit antennas at the BS becomes large
    to evaluate the performance of our proposed 2PCE strategy.  The analysis shows that the proposed strategy is able to outperform the existing 3PCE strategy with the same channel training overhead.
  Finally, extensive numerical results are presented to validate the effectiveness of the proposed 2PCE strategy under various numbers of BS antennas, IRS reflecting elements and users, and we show that the performance of the proposed 2PCE strategy is better than that of the 3PCE strategy, even when the number of BS antennas is not large. The impacts of various system parameters, such as the channel training signal-to-noise ratio (SNR), the path-loss exponents, the channel Rician factors and the correlation coefficients   on the channel estimation performance are investigated and useful insights are drawn.

The rest of this paper is organized as follows.
Section II presents  the considered IRS-aided uplink multiuser  SIMO system model. Section III  introduces the details of the proposed 2PCE strategy  under the ideal case without receive noise at the BS. Theoretical analysis on the asymptotic MSE of the proposed 2PCE strategy is presented in Section IV  and comparison with the existing 3PCE strategy is provided. Numerical results are presented in Section V and finally, Section VI concludes this paper. 

\textit{Notation:} Scalars, vectors and matrices are respectively denoted by lower (upper) case, boldface lower case and boldface upper case letters. The symbol $[\mathbf{X}]_{n,m}$  denotes the $(n,m)$-th entry of the matrix $\mathbf{X}$ while $[\mathbf{x}]_n$ represents the $n$-th entry of the vector $\mathbf{x}$. { The symbols $[\cdot,\cdot]$ and $[\cdot;\cdot]$ represent  the  column and row concentrations of  two scalars/vectors/matrices.}  For a matrix $\mathbf{X}$ of arbitrary size, $\mathbf{X}^T$, $\mathbf{X}^*$, $\mathbf{X}^H$ and $\mathbf{X}^{\dagger}$  denote the transpose, conjugate, conjugate transpose and pseudo-inverse    of $\mathbf{X}$, respectively. { For a square matrix $\mathbf{Y}$, $\text{tr}(\mathbf{Y})$ represents the trace of $\mathbf{Y}$, and  $\mathbf{Y}^{-1}$ denotes the inverse of $\mathbf{Y}$ if $\mathbf{Y}$ is full-rank. } 
The symbol $||\cdot||$ denotes the Euclidean norm of a complex vector, and $|\cdot|$ is the absolute value of a complex scalar. The symbol $\text{diag}(x_1,\cdots, x_N)$ denotes a diagonal matrix whose diagonal
elements are set as $x_1,\cdots,x_N$. ${\mathbb{C}}^{m\times n}$  denotes the space of $m \times n$ complex matrices. $\lceil x\rceil$ and $\lfloor x\rfloor$ represent the minimum integer which is larger than  $x$ and the maximum integer that is smaller than $x$, respectively. The symbol $\%$ represents the remainder operation, i.e., $x_1\% x_2 = x_1 - x_2\lfloor \frac{x_1}{x_2}\rfloor$. $\mathbf{I}_M$, $\mathbf{0}_{M\times N}$ and $\mathbf{1}_{M\times N}$ denote the $M \times M$ identity matrix,  $M \times N$ all-zero matrix and  $M \times N$ all-one matrix, respectively. Finally, we define the complex normal distribution as $\mathcal{C}\mathcal{N}(\mu,{\sigma}^2)$ with mean $\mu$ and variance ${\sigma}^2$.

\vspace{-0.5cm}\section{System Model}\label{SecII}
\begin{figure}[b]
\vspace{-1.0cm}
\setlength{\belowcaptionskip}{-0.3cm}
\renewcommand{\captionfont}{\small}
\centering
\includegraphics[scale=.53]{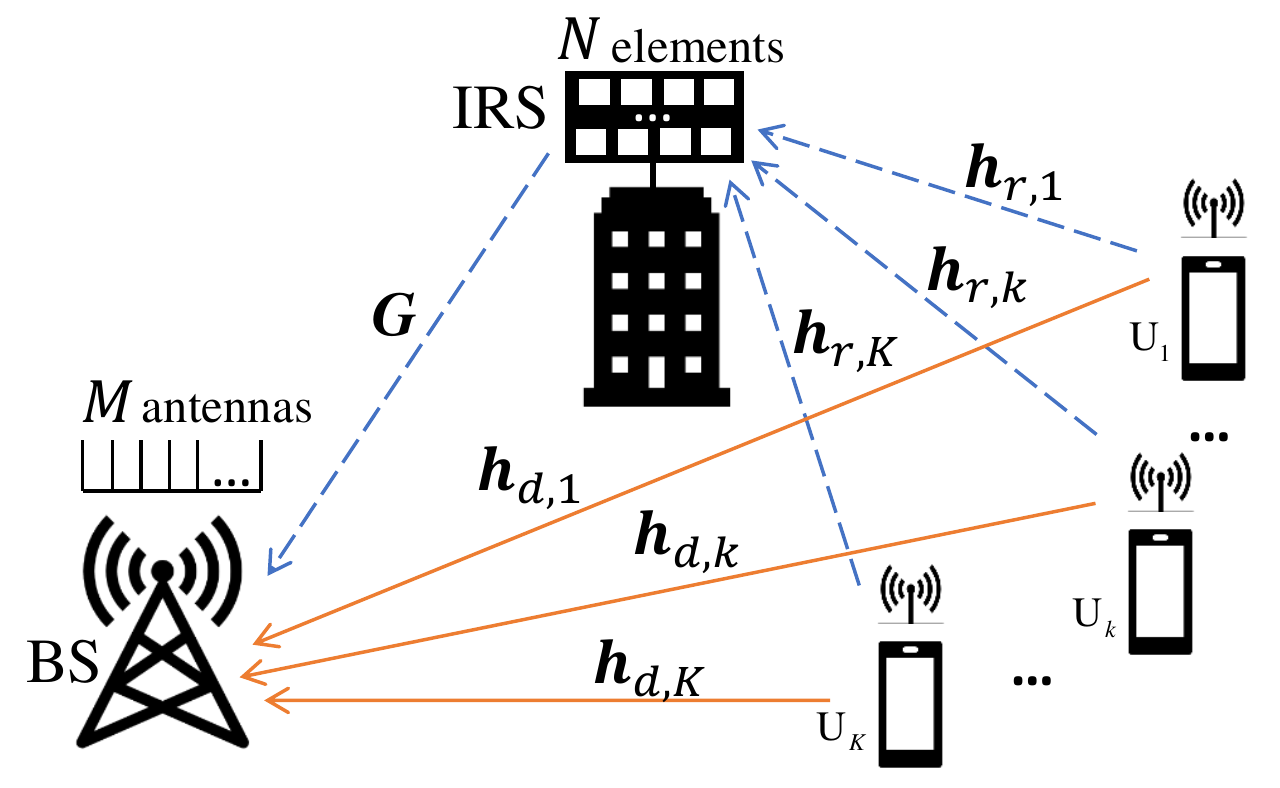}
\caption{An IRS-aided multiuser SIMO communication system.}
\label{system}
\normalsize
\end{figure}
We consider an  IRS-aided uplink multiuser  SIMO system as shown in Fig. 1, where $K$ single-antenna users, denoted by $\{U_{1}, \cdots, U_{K}\}$, simultaneously communicate with a BS, and an IRS is deployed to enhance the communication performance. The BS is equipped with a uniform linear array (ULA) with $M$ antennas, and the IRS is equipped with an $N_y \times N_z$ uniform planar array (UPA) which contains $N =N_yN_z $ passive reflecting elements. Let $\mathbf{h}_{d,k} \in \mathbb{C}^{M \times 1}$ and $\mathbf{h}_{r,k} \in \mathbb{C}^{N \times 1}$ denote the baseband equivalent channels from $U_{k}$ to the BS and the IRS, respectively. The symbol ${\mathbf{G}} \in \mathbb{C}^{M \times N}$  represents the  channel from the IRS to the BS with $\mathbf{g}_n $ as the $n$-th column of ${\mathbf{G}} $.

In this work,  a quasi-static block-fading channel is assumed and all the channels $\mathbf{h}_{d,k}$'s, $\mathbf{h}_{r,k}$'s, $\mathbf{G}$ remain approximately constant over a coherence interval of length $T$ symbols. Since  both line-of-sight (LoS) and non-LoS (NLoS) components may exist in practical channels  due to the insufficient angular spread of the scattering environment and closely
spaced antennas/reflecting elements, $\mathbf{h}_{d,k}$  can be modeled as  the general spatially correlated Rician fading channel \cite{McKay2005} as
\vspace{-0.3cm}
 \begin{equation}\label{eq37}
{\mathbf{h}}_{d,k} = \sqrt{\frac{\beta^{\text{UB}}}{1+\beta^{\text{UB}}}}{\bar{\mathbf{z}}}_{d,k}+\sqrt{\frac{1}{1+\beta^{\text{UB}}}}{\bm{\Phi}}_d^{\frac{1}{2}}{{\mathbf{z}}}_{d,k},
 \end{equation}

 \vspace{-0.2cm}\noindent
 where $\beta^{\text{UB}}$ denotes the  Rician factor; ${\bar{\mathbf{z}}}_{d,k}\in {\mathbb{C}}^{M\times 1}$  denotes LoS  component in ${\mathbf{h}}_{d,k}$; $[{{\mathbf{z}}}_{d,k}]_{m}\sim \mathcal{C} \mathcal{N}\left(0, l_{k,m}^{\text{UB}} \right)$ follows the independent and identically distributed (i.i.d.) Rayleigh fading channel model with $\l_{k,m}^{\text{UB}}$ denoting the distance-dependent path loss of the link from $U_k$ to the $m$-th BS antenna; ${\bm{\Phi}}_d$ is the BS receive correlation matrix.
 Similarly, the user-IRS and IRS-BS channels  can be modeled as follows:
 \vspace{-0.3cm}
  \begin{equation}\label{eq7}
{\mathbf{h}}_{r,k} = \sqrt{\frac{\beta^{\text{UI}}}{1+\beta^{\text{UI}}}}{\bar{\mathbf{z}}}_{r,k}+\sqrt{\frac{1}{1+\beta^{\text{UI}}}}{\bm{\Phi}}_{r,k}^{\frac{1}{2}}{{\mathbf{z}}}_{r,k},
 \;\;
 {\mathbf{G}} = \sqrt{\frac{\beta^{\text{IB}}}{1+\beta^{\text{IB}}}}\bar{\mathbf{F}}+\sqrt{\frac{1}{1+\beta^{\text{IB}}}}{\bm{\Phi}}_d^{\frac{1}{2}}{\mathbf{F}}{\bm{\Phi}}_r^{\frac{1}{2}},
 \end{equation}

 \vspace{-0.2cm}\noindent
where $\beta^{\text{UI}}$ and $\beta^{\text{IB}}$  denote the corresponding  Rician factors;  ${\bm{\Phi}}_{r,k}$  represents the correlation matrix  between the IRS and $U_k$, and ${\bm{\Phi}}_r$ is the IRS transmit correlation matrix; $[{{\mathbf{z}}}_{r,k}]_{n}\sim \mathcal{C} \mathcal{N}\left(\mathbf{0},l^{\text{UI}}_{k,n}\right)$ and $[{\mathbf{F}}]_{m,n}\sim \mathcal{C} \mathcal{N}\left(\mathbf{0},l^{\text{IB}}_{m,n}\right)$  denote the NLoS Rayleigh fading components with $l^{\text{UI}}_{k,n}$ and $l^{\text{IB}}_{m,n}$  denoting the corresponding path losses, while ${\bar{\mathbf{z}}}_{r,k}$ and $\bar{\mathbf{F}}$  represent the LoS components.

More specifically,  the LoS components of the user-BS, user-IRS and IRS-BS channels, i.e., ${\bar{\mathbf{z}}}_{d,k}$, ${\bar{\mathbf{z}}}_{r,k}$ and $\bar{\mathbf{F}}$,  are respectively modeled as:
 \vspace{-0.3cm}\begin{equation}\label{eq4}
[{\bar{\mathbf{z}}}_{d,k}]_m = \sqrt{l^{\text{UB}}_{k,m}}{f}_{L}({\vartheta}_{k}^{\text{UB}}),\;\;
[{\bar{\mathbf{z}}}_{r,k}]_n = \sqrt{l^{\text{UI}}_{k,n}}{f}_{P}(\varphi^{\text{UI}}_{k},\psi^{\text{UI}}_{k}),\;\;
[{\bar{\mathbf{F}}}]_{m,n} = \sqrt{l^{\text{IB}}_{m,n}}{f}_{L}(\vartheta^{\text{IB}}){f}_{P}(\varphi^{\text{IB}},\psi^{\text{IB}}),
 \end{equation}

 \vspace{-0.2cm}\noindent
where ${f}_{P}(\varphi,\psi) =e^{{{j 2 \pi d_{\text{I}}\big((n-\lfloor {n}/{ N_{x}}\rfloor N_{x}) \sin ({\varphi}) \cos ({\psi})+\lfloor {n}/{ N_{x}}\rfloor \sin ({\varphi}) \sin ({\psi})\big) } / {\lambda}}}$, ${f}_{L}(\vartheta) = e^{{{j 2 \pi(m-1) d_{\text{B}} \sin ({\vartheta})}/{\lambda}}}$,
$\lambda$ is the wavelength, $d_B$ denotes the antenna spacing at the BS, $d_I$ denotes the  reflecting element spacing at the IRS; ${\vartheta}_{k}^{\text{UB}}$ and ${\vartheta}^{\text{IB}}$ denote the angles-of-arrival (AoAs) from $U_k$ to the BS and that from the IRS to the BS, respectively;  ${\psi}^{\text{IB}}$ and ${\varphi}^{\text{IB}}$ are  the azimuth and elevation angles-of-departure (AoDs) from the IRS  to the BS, respectively; ${\psi}^{\text{UI}}_k$ and ${\varphi}^{\text{UI}}_{k}$ represent the azimuth and elevation AoAs from $U_k$ to the IRS, respectively.
Furthermore, according to \cite{Loyka2001,YiWei2019}, the BS receive correlation matrix can be modeled as
 \vspace{-0.3cm}\begin{equation}\label{eq43}
 {\bm{\Phi}}_d(i, j)=\left\{\begin{array}{ll}
(r_{ d})^{j-i}, &\text { if } i \leq j, \\
{\bm{\Phi}}_d^*(j, i), &\text { otherwise },
\end{array}\right.
 \end{equation}

 \vspace{-0.2cm}\noindent
where $r_d \in [0,1]$ is the correlation coefficient. Besides, ${\bm{\Phi}}_r$ (${\bm{\Phi}}_{r,k}$) can be modeled as ${\bm{\Phi}}_r = {\bm{\Phi}}_r^h \otimes {\bm{\Phi}}_r^v$ (${\bm{\Phi}}_{r,k} ={\bm{\Phi}}_{r,k}^h \otimes {\bm{\Phi}}_{r,k}^v$) according to \cite{Choi2014},
where ${\bm{\Phi}}_r^h$ (${\bm{\Phi}}_{r,k}^h$)
 and ${\bm{\Phi}}_r^v$ (${\bm{\Phi}}_{r,k}^v$) denote the spatial correlation matrices of the horizontal and vertical domains,
respectively, and they are similarly defined as in \eqref{eq43} with $r_r$ ($r_{r,k}$) denoting the correlation coefficient.

In the considered IRS-aided uplink multiuser SIMO  system, each reflecting element at the IRS is able to   induce an independent phase-shift change of the incident signal sent by the users.  Thus, by properly adjusting the IRS reflection
coefficients at different reflecting elements, a preferable wireless signal propagation environment between the transmitters and receiver can be created to enhance the
communication performance.
Let $\theta_{n,i}$ denote the reflection coefficient of the $n$-th reflecting element in  time slot $i$, which satisfies
$|\theta_{n, i}| = 1$ if the $n$-th reflecting element    is on and $|\theta_{n, i}| = 0$ otherwise.
It is shown that the phase of the incident signal on the $n$-th reflecting element can only be adjusted when the reflecting element is switched on.
The received signal at the BS in time slot $i, i \in [1,T]$,
is given by
 \vspace{-0.3cm}\begin{equation}\label{eqa1}
\mathbf{y}^{(i)} =\sum_{k=1}^{K} \mathbf{h}_{d,k} \sqrt{p_k} s_{k}^{(i)}+\sum_{k=1}^{K} \sum_{n=1}^{N} \theta_{n, i} [\mathbf{h}_{r,k}]_n \mathbf{g}_{n} \sqrt{p_k} s_{k}^{(i)}+\mathbf{n}^{(i)},
 \end{equation}

 \vspace{-0.3cm}\noindent
where $s_k$ and $p_k$ denote the transmit signal and transmit power of $U_k$, respectively, $\mathbf{n}^{(i)} \sim \mathcal{C} \mathcal{N}\left(\mathbf{0}, \sigma^{2} \mathbf{I}\right)$ denotes the additive white Gaussian noise (AWGN) at the BS in time slot $i$. 
Define the reflected channel from $U_k$ to the BS via the IRS as
 \vspace{-0.3cm}\begin{equation}\label{A-11}
{\mathbf{H}}_k \triangleq \mathbf{G}\text{diag}({\mathbf{h}}_{r,k})  \in {\mathbb{C}}^{M\times N},
 \end{equation}

 \vspace{-0.3cm}\noindent
  and let $\mathbf{h}_{k,n}$ denote the $n$-th column of $\mathbf{H}_{k}$, then,
the received signal in \eqref{eqa1} can be   equivalently rewritten as\footnote{In this work, we assume for simplicity that the transmit power of each user is set to the maximum value $p$ to maximize the channel training SNR, i.e., $p_k=p, \forall k \in [1,K]$.}
 \vspace{-0.3cm}\begin{equation}\label{eq2}
\mathbf{y}^{(i)} =\sum_{k=1}^{K}\big(\mathbf{h}_{d,k}+{\mathbf{H}}_k\bm{\theta}_{i}) \sqrt{p} s_{k}^{(i)}+\mathbf{n}^{(i)},
 \end{equation}

 \vspace{-0.3cm}\noindent
where $\bm{\theta_{i}} = [\theta_{1,i};\cdots;\theta_{N,i}]$.
We can observe from \eqref{eq2} that in order to recover the transmit signal $
\{s^{(i)}_k\}_{k=1}^K$ from the received signal $\mathbf{y}^{(i)}$, all the direct channels $\{\mathbf{h}_{d,k}\}_{k=1}^K$ and reflected channels $\{\mathbf{H}_{k}\}_{k=1}^K$ need to be obtained.

Generally, each coherence block of length $T$ time slots is divided into the channel estimation stage (consists of $\tau<T$ time slots) and data transmission stage (consists of $T-\tau$
time slots).
Let ${a}_{k,i}$ denote the pilot symbol of $U_k$ in time slot $i$  which satisfies $|a_{k,i}|\in \{0,1\},i\in [1,T]$.
Since  the IRS is not capable
of transmitting or receiving  pilot symbols, the BS is expected to estimate $KM + KMN$ unknown coefficients included in $\{\mathbf{h}_{d,k},\mathbf{H}_{k}\}_{k=1}^K$ based on the overall received signal $\mathbf{Y}\triangleq[\mathbf{y}^{(1)}, \cdots, \mathbf{y}^{(\tau)}] \in \mathbb{C}^{M \times \tau}$, the known user pilot symbol $\{a_{k, i}\}_{k=1}^K,i\in [1,\tau]$ and the IRS reflection coefficients $\{\theta_{n,i}\}_{n=1}^N,i\in [1,\tau]$. {When the numbers of BS antennas,  IRS reflecting elements and/or  users are large,  a great amount of channel training overhead is required to estimate such  a large number of channel coefficients, which will reduce  the sum rate of the considered IRS-aided communication system. } Motivated by the 3PCE strategy in \cite{arXiv1912}, we propose a new channel estimation strategy to exploit the strong correlation among $\mathbf{H}_k$'s for channel training overhead reduction and in the meantime, alleviate the error propagation issue by using fewer channel estimation phases. The robust beamforming design for data transmission with given  estimated channels and   channel error distributions will be left to our future work.
%

\vspace{-0.4cm}\section{Proposed 2PCE Strategy}\label{SIII}
\newtheorem{proposition}{\bf Proposition}
\newtheorem{lemma}{\bf Lemma}
In this section, a novel 2PCE strategy for the considered IRS-aided multiuser  SIMO  system is proposed.
The main idea is to alleviate the error propagation issue by reducing the channel estimation phases, which is essentially achieved by estimating the direct and reflected channels associated with each user simultaneously.
Specifically, the 3PCE strategy in \cite{arXiv1912} works as follows: 1) in the first phase, all the direct channels  are estimated by switching off all the reflecting elements; 2) in the second phase, the IRS is switched
on and only one typical user is allowed to transmit non-zero pilot symbols, then the
reflected channel associated with this typical user is estimated based on its estimated direct channel in the first phase; 3) in the last phase, due to the fact that the  reflected channels associated with the other users are scaled versions of the  reflected channel associated with the typical user, the scaling factors (rather than the whole reflected channels)  are estimated  based on the  estimated  reflected channel associated with the typical user and  the direct channels.    We can observe from the above 3PCE strategy that the estimated direct channels obtained in the first phase are required in the next two phases, which means that the estimation errors of the direct channels will affect
 the estimation accuracy of all the reflected channels,
 especially when the channel estimation in the first phase is not very accurate.\footnote{Note that similar error propagation issues also exist in other channel estimation strategies, e.g., \cite{arXiv01301,Mishra2019,9039554}, etc.}
 Besides, estimation errors of the  reflected channel associated with the typical user in the second phase will also deteriorate  the estimation performance of the  reflected channels associated with other users.
 Therefore, the performance of the state-of-the-art 3PCE strategy in \cite{arXiv1912} is limited by this error propagations issue and it can be seen that the channel estimation performance of the second and third phases are critically dependent on those in the first and second phases.

\begin{figure}[t]
\vspace{-0.5cm}
\setlength{\belowcaptionskip}{-0.9cm}
\renewcommand{\captionfont}{\small}
\centering
\includegraphics[scale=.35]{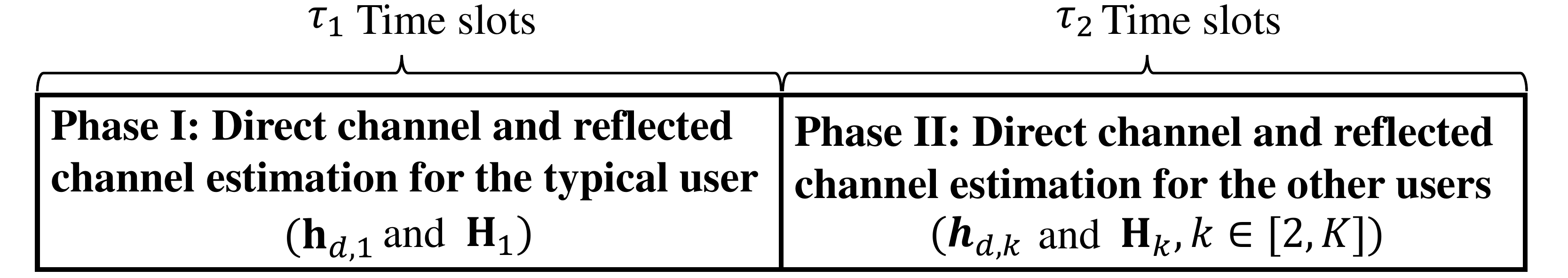}
\caption{Proposed 2PCE strategy. }
\label{chaEst}
\normalsize
\end{figure}

In the following, we present the proposed  2PCE  strategy to alleviate the abovementioned  error propagation issue.
As shown in Fig. \ref{chaEst}, the proposed strategy is divided into two phases:
{ \begin{itemize}
\item In Phase I which consists of $\tau_1$ time slots, the direct and reflected channels associated with a typical user (denoted by $U_1$ for convenience), i.e.,  $\mathbf{h}_{d,1}$ and $\mathbf{H}_1$, are estimated simultaneously;
\item In Phase II which consists of $\tau_2$ time slots, the direct and reflected channels associated with $U_k$, $k\in [2,K]$, are estimated simultaneously based on the estimated  reflected channel associated with the typical user by exploiting the correlation among  $\mathbf{H}_k$'s and properly designing the pilot symbols of the users and the reflection patterns.
\end{itemize}}
\noindent Note that in the proposed 2PCE strategy, the reflected channel associated with the typical user is estimated in the first phase and it does not depend on the estimation accuracy of the direct channels, therefore the estimation accuracy of $\mathbf{H}_1$ can be improved without increasing the channel training overhead.
 Moreover,
 the direct and reflected channels associated with each user are estimated jointly, hence the estimation errors of the direct channels will not propagate to the estimation process of the reflected channels.


For a better illustration of the overall channel estimation process of the 2PCE strategy and the required channel  training  overhead (i.e., the specific numbers of $\tau_1$ and $\tau_2$), we consider the ideal case without receive noise
at the BS in the rest of this section. How to implement our proposed 2PCE framework in the practical case with noise and the detailed analysis of the asymptotic MSE for estimating the channels will be presented in the next section.
\vspace{-0.5cm}\subsection{Phase I: Direct and Reflected Channels Estimation for Typical User}
In Phase I, all the reflecting elements are switched on and only the typical user $U_1$ transmits non-zero pilots to the BS,
such  that the direct and  reflected channels associated with $U_1$ can be estimated.
Accordingly,  the received signal at the BS in time slot $i$ is given by
 \vspace{-0.3cm}\begin{equation}\label{eq3}
\mathbf{y}^{(i)}={\mathbf{H}}_1\boldsymbol{\theta}_{i} \sqrt{p} a_{1, i}+\mathbf{h}_{d,1} \sqrt{p} a_{1, i}, \;\; i\in [1,\tau_1],
 \end{equation}

 \vspace{-0.3cm}\noindent
which can be viewed as a linear system  of $M$ equations. Since there are $M(N+1)$ unknown channel coefficients included in ${\mathbf{h}}_{d,1}$ and $ {\mathbf{H}}_1$, at least  $M(N+1)$ linear equations of these unknowns are required to estimate $\{\mathbf{h}_{d,1}, \mathbf{H}_1\}$, thus at least $ N+1$ time slots are needed in this phase.
By defining $
{\tau}_1 = N+1
$ and $a_{k,i} = 1, i\in [1,\tau_1]$,
the overall received signal at the BS in  Phase I can be written as
 \vspace{-0.3cm}\begin{equation}\label{eq44}
{\mathbf{Y}}^{\text{I}} \triangleq  [{\mathbf{y}}^{(1)},\cdots,{\mathbf{y}}^{(N+1)}] =\sqrt{p}{\mathbf{X}}^{\text{I}}{\mathbf{V}}^{\text{I}},
 \end{equation}

 \vspace{-0.25cm}\noindent
where
${\mathbf{X}}^{\text{I}} \triangleq [{\mathbf{h}}_{d,1},{\mathbf{H}}_1] \in {\mathbb{C}}^{M\times (N+1)}$  denotes the composite channel matrix  associated with $U_1$,  and ${\mathbf{V}}^{\text{I}} \in {\mathbb{C}}^{(N+1)\times (N+1)}$ represents
 the reflection pattern matrix
 the $i$-th column of which is given by $[1;\bm{\theta}_{i}]$. To minimize the variance of the channel estimation error,  $\mathbf{V}^{\text{I}}$  should be an  orthogonal matrix and
each of its entry should satisfy the uni-modulus constraint \cite{Jensen2019}. In this paper, similar to \cite{Jensen2019,OFDM2020,MingMin2020},   we set  $\mathbf{V}^{\text{I}}$   as an $(N+1)$-dimension DFT matrix which satisfies $({\mathbf{V}}^{\text{I}})^{-1}= \frac{({\mathbf{V}}^{\text{I}})^{H}}{N+1}$. Hence,
 the estimated composite channel can be expressed as
 \vspace{-0.3cm}\begin{equation}\label{eq50}
\hat{{\mathbf{X}}^{\text{I}}} \triangleq [\hat{{\mathbf{h}}}_{d,1},\hat{{\mathbf{H}}}_1] = \frac{{\mathbf{Y}}^{\text{I}}({\mathbf{V}}^{\text{I}})^{H}}{\sqrt{p}(N+1)}.
 \end{equation}
\vspace{-1.5cm}\subsection{Phase II: Direct and Reflected Channels Estimation for Other Users}\label{III-B}
In Phase II, the other users, i.e., $\{U_k\}_{k=2}^K$,  transmit non-zero pilot symbols to the BS  to estimate the direct and reflected channels, i.e.,  $\{\mathbf{h}_{d,k},\mathbf{H}_k  \}_{k=2}^K$. The received signal at the BS in time slot $i$  can be written as
 \vspace{-0.35cm}\begin{equation}\label{A-14}
\mathbf{y}^{(i)}=\sum_{k=2}^{K} {\mathbf{H}}_k\boldsymbol{\theta}_{i}\sqrt{p} a_{k, i}+\sum_{k=2}^{K} \mathbf{h}_{d,k} \sqrt{p} a_{k, i}, \;\; i \in [\tau_1 + 1,\tau_1 + \tau_2].
 \end{equation}

 \vspace{-0.3cm}\noindent
 Intuitively, if the correlations among $\mathbf{H}_k$'s is not exploited,  at least $(K-1)(N+1)$ time slots are required to estimate the $(K-1)(MN+M)$ channel coefficients in $\{\mathbf{h}_{d,k},\mathbf{H}_k  \}_{k=2}^K$. When the numbers of BS antennas, reflecting elements and users are large, the channel  training overhead required in Phase II would be overwhelming, which leads to low user transmission rate since the time left for data transmission is limited in this case.
To overcome this difficulty,  we adopt the strategy in \cite{arXiv1912} and propose to express the reflected channel associated with  $U_k$ as a scaled version of $\mathbf{H}_1$, which is shown as follows:
 \vspace{-0.45cm}\begin{equation}\label{A-10}
\mathbf{H}_k = \mathbf{H}_1 \text{diag}(\bm{\mu}_k),
 \end{equation}

 \vspace{-0.35cm}\noindent
 where $\bm{\mu}_k \in \mathbb{C}^{N\times 1}$ represents the scaling vector of $\mathbf{H}_k$ and the $n$-th entry of $\bm{\mu}_k$, i.e., $\mu_{k,n}={[\mathbf{h}_{r,k}]_n}/{[\mathbf{h}_{r,1}]_n}$, represents the scaling factor associated with the $n$-th reflecting element. Utilizing  the relationship in \eqref{A-10}, $\{\mathbf{H}_k\}_{k=2}^K$ can be estimated  based on the knowledge of  $\{\bm{\mu}_k\}_{k=2}^K$ and $\mathbf{H}_1$. Thus, we can estimate $\big\{\{\mathbf{h}_{d,k}\}_{k=1}^K,\mathbf{H}_1, \{\bm{\mu}_k\}_{k=2}^K\big\}$  instead of $\big\{ \{\mathbf{h}_{d,k}\}_{k=1}^K,\{\mathbf{H}_k\}_{k=1}^K \big\}$, and  the number of unknown channel  coefficients to be estimated can be significantly reduced. As a result,
 the received signal in \eqref{A-14} can be equivalently rewritten as
 \vspace{-0.3cm}\begin{equation}\label{eq36}
 {\mathbf{y}}^{(i)}=\sqrt{p}\sum_{k=2}^K a_{k,i}({\mathbf{h}}_{d,k}+{\mathbf{H}}_{1}\text{diag}({\boldsymbol{\theta}}_i){\bm{\mu}}_k), \;\; i \in [{\tau}_1 + 1,{\tau}_1 + {\tau}_2].
\end{equation}
\vspace{-0.7cm}

Let us define
${\mathbf{x}}^{\text{II}} \triangleq [{\bm{\mu}}_2;\cdots;{\bm{\mu}}_K;{\mathbf{h}}_{d,2};\cdots;{\mathbf{h}}_{d,K}]$, $ {\mathbf{y}}^{\text{II}} \triangleq [ {\mathbf{y}}^{(\tau_1+1)};\cdots;{\mathbf{y}}^{(\tau_1+\tau_2)}]$ and
 \vspace{-0.3cm}\begin{equation}\label{WII}
{\mathbf{W}}^{\text{II}} \triangleq \left[\begin{array}{cccccc}
a_{2,\tau_1+1}{\mathbf{H}}_{1}\text{diag}({\boldsymbol{\theta}}_{\tau_1+1})&\cdots &a_{K,\tau_1+1}{\mathbf{H}}_{1}\text{diag}({\boldsymbol{\theta}}_{\tau_1+1}) &a_{2,\tau_1+1}{{\mathbf{I}}_M} &\cdots &a_{K,\tau_1+1}{{\mathbf{I}}_M}\\
\vdots & \ddots & \vdots & \vdots & \ddots & \vdots \\
a_{2,\tau_1+\tau_2}{\mathbf{H}}_{1}\text{diag}({\boldsymbol{\theta}}_{\tau_1+\tau_2})&\cdots &a_{K,\tau_1+\tau_2}{\mathbf{H}}_{1}\text{diag}({\boldsymbol{\theta}}_{\tau_1+\tau_2}) &a_{2,\tau_1+\tau_2}{{\mathbf{I}}_M} &\cdots &a_{K,\tau_1+\tau_2}{{\mathbf{I}}_M}
\end{array}\right],
\end{equation}
then the overall received signal at the BS in Phase II is given by
 \vspace{-0.3cm}\begin{equation}\label{A17}
 {\mathbf{y}}^{\text{II}}=\sqrt{p}{\mathbf{W}}^{\text{II}}{\mathbf{x}}^{\text{II}},
 \end{equation}

 \vspace{-0.25cm}\noindent
which is  a linear system with $\tau_2M$ equations and $(K-1)(M+N)$ unknowns.

To show how to design the pilot symbols $a_{k,i}$'s 
and the reflection patterns $\bm{\theta}_{i}$'s such that \eqref{WII} can be solved efficiently, we resort to the following lemma which indicates the necessary and sufficient  condition for a linear system of  equations to have a unique solution.
\vspace{-0.3cm}
\begin{lemma}\label{linearLemma}\rm
\cite{NA1993} For any  linear system of  equations $\mathbf{Wx}=\mathbf{y}$ with $\mathbf{W}\in {\mathbb{C}}^{d_1\times d_2},\mathbf{x}\in {\mathbb{C}}^{d_2\times 1}$ and ${\mathbf{y}}\in {\mathbb{C}}^{d_1\times 1}$, there exists a unique solution  { if and only if} the matrix $\mathbf{W}$ satisfies: 1) $d_1\ge d_2$, 2) $\text{rank}({\mathbf{W}})=d_2$.
 \end{lemma}
 \vspace{-0.2cm}
 Based on Lemma \ref{linearLemma}, we observe that $a_{k,i}$'s and $\bm{\theta}_{i}$'s should be carefully designed to satisfy $\text{rank}({\mathbf{W}}^{\text{II}}) = (K-1)(M+N)$  and $\tau_2M\ge (K-1)(M+N)$ such that $\mathbf{x}^{\text{II}}$ can be uniquely determined by \eqref{A17}.
 { Let us consider the following two cases for details.}

\subsubsection{$M\ge N$}
{In this case, all the users except $U_1$ are allowed to  transmit non-zero pilots simultaneously, and orthogonal pilot sequences are employed to eliminate the inter-user interference.  In order to obtain the unique solution of \eqref{A17}, i.e., $\{\bm{\mu}_{k},\mathbf{h}_{d,k}\}_{k=2}^K$, we set $\tau_2 = 2(K-1)$ and design the following pilot sequence ${\mathbf{A}}^{\text{II}}$ and reflection patterns $\{{\boldsymbol{\theta}}_{i}\}_{i=\tau_1+1}^{\tau_1+\tau_2}$ to meet the requirements in  Lemma \ref{linearLemma}:
 \vspace{-0.3cm}\begin{equation}\label{A14}
{\mathbf{A}}^{\text{II}}
\triangleq[{\mathbf{a}}_2^{\text{II}},\cdots,{\mathbf{a}}_K^{\text{II}}] =[\mathbf{D}_{K-1};\mathbf{D}_{K-1}],\;\;
{\boldsymbol{\theta}}_{i} = \left\{\begin{array}{ll}
 \;\mathbf{1}_{N\times 1}, &\text{ if } i\in [\tau_1+1,\tau_1+K-1],\\
 \;-\mathbf{1}_{N\times 1}, &\text{ otherwise,}
\end{array}\right.
 \end{equation}

 \vspace{-0.25cm}\noindent
where $\mathbf{a}^{\text{II}}_k = [a_{k,\tau_1};\cdots; a_{k,\tau_1+\tau_2}]$ and $\mathbf{D}_{K-1}$ denotes the $(K-1)$-dimension DFT matrix.
According to \eqref{A14}, $\mathbf{x}^{\text{II}}$ can be obtained by multiplying $\mathbf{y}^{\text{II}}$ { on the right hand side (RHS) with} the pseudo-inverse of $\mathbf{W}^{\text{II}}$.  } However, the direct calculation of $(\mathbf{W}^{\text{II}})^{\dagger}$ can be very computationally intensive due to the large dimension of $\mathbf{W}^{\text{II}}$.
To address this problem, we first design an auxiliary matrix
\vspace{-0.2cm}\begin{equation}\label{SIGMA}
{\bm{\Sigma}}_k
\triangleq  \left[\begin{array}{cccccc}
a_{k,\tau_1+1}^{*}{\mathbf{I}}_M  & \cdots & a_{k,\tau_1+K-1}^{*}{\mathbf{I}}_M & {\mathbf{0}}_{M\times M}  & \cdots & {\mathbf{0}}_{M\times M}\\
{\mathbf{0}}_{M\times M} & \cdots & {\mathbf{0}}_{M\times M} & a_{k,\tau_1+K}^{*}{\mathbf{I}}_M &  \cdots & a_{k,\tau_1+2(K-1)}^{*}{\mathbf{I}}_M
\end{array}\right] \in {\mathbb{C}}^{2M\times (N+M)(K-1)},
\end{equation}

\vspace{-0.2cm} \noindent
then by left-multiplying  $\mathbf{y}^{\text{II}}$ with  ${\bm{\Sigma}}_k,k \in [2,K]$,
 \eqref{A17} can be decomposed into $K-1$ smaller linear systems of equations, which are shown as follows:
 \vspace{-0.3cm}\begin{equation}\label{proof1a}
 {\bm{\Sigma}}_k {\mathbf{y}}^{\text{II}}=\sqrt{p}(K-1){\mathbf{V}}^{\text{II}} [{\bm{\mu}}_k;
 {\mathbf{h}}_{d,k}],\;\; k \in [2,K],
 \end{equation}

 \vspace{-0.3cm}\noindent
where ${\mathbf{V}}^{\text{II}} \triangleq [{\mathbf{H}}_1, {{\mathbf{I}}_M}; -{\mathbf{H}}_1, {{\mathbf{I}}_M}]$.
As can be observed from the channel model in \eqref{eq7} and \eqref{A-11},
the columns of $\mathbf{H}_1$  are linearly independent with probability one, thus the matrix ${\mathbf{V}}^{\text{II}}$ is of full column rank. Besides, the number of rows in ${\mathbf{V}}^{\text{II}}$ is larger than its number of columns when $M\ge N$. Therefore, for each $k \in [2,K]$, there exists a unique solution to \eqref{proof1a}, and $\{{\bm{\mu}}_k,{\mathbf{h}}_{d,k}\}_{k=2}^K$ can be perfectly estimated  as
 \vspace{-0.4cm}\begin{equation}\label{B-1}
[{\hat{\bm{\mu}}}_k;
 {\hat{\mathbf{h}}}_{d,k}] = \frac{1}{\sqrt{p}(K-1)}({\mathbf{V}}^{\text{II}})^{\dagger}{\bm{\Sigma}}_k {\mathbf{y}}^{\text{II}},\;\; k \in [2,K].
\end{equation}
 \vspace{-0.8cm}\subsubsection{$M\textless N$}
 In this case, in order for \eqref{A17} to have  a unique solution, at least $\lceil\frac{(K-1)(M+N)}{M} \rceil = K-1+ \lceil\frac{ (K-1)N}{M}\rceil$ time slots are needed to meet the first requirement in Lemma \ref{linearLemma}.
 However, compared with the case of $M\ge N$, it is much more challenging in this case to design a matrix $\mathbf{W}^{\text{II}}$ that satisfies the full-rank requirement when all the users are active. For ease of pilot sequence and reflection pattern design, we only allow one or  a few users to transmit non-zero pilot symbols in each time slot, as inspired by \cite{arXiv1912}. Our  main idea to achieve the minimum channel training overhead, i.e., $\tau_2  = K-1+ \lceil\frac{(K-1)N}{M}\rceil$, is to allow the users to share some of the $\tau_2$ time slots with a selected number of reflecting elements switched on.
More specifically, the proposed strategy in this case can be further divided into two subphases, i.e., Phase II-A and Phase II-B. Let us define $\gamma \triangleq \lfloor \frac{M+N}{M} \rfloor $ and $\delta \triangleq N - (\gamma -1)M$, then
  Phase II-A (which consists of $(K-1)\gamma$ time slots) is divided into $K-1$ separate non-overlapping time intervals each consists of $\gamma$ time slots and each user is allocated a separate time interval. In this subphase, each user
 transmits an all-one pilot sequence with some pre-selected reflecting elements at the IRS switched on, such that the corresponding  direct channel and $M(\gamma-1)$ scaling factors can be estimated.   In Phase II-B which consists of $\lceil\frac{(K-1)\delta}{M}\rceil$ time slots, more than one users are activated and the rest reflecting elements that are not selected in Phase II-A are switched on. Thereby, the rest $(K-1)\delta$ scaling factors associated with each user can be obtained based on the estimated $\mathbf{H}_1$ obtained in Phase I as well as the estimated direct channels and  $M(\gamma-1)(K-1)$ scaling factors obtained in Phase II-A.

 Let $\chi_k^{\text{A}}$ and $\chi_k^{\text{B}}$ denote the index sets of the scaling factors which are estimated  in Phase II-A and Phase II-B, respectively, we let
 \vspace{-0.3cm}\begin{equation}\label{eq55}
\chi_k^{\text{B}} =\big \{\big((k-2)\delta+1\big)\;\% \; N,\cdots, \big((k-1)\delta\big)\;\%\; N\big\}, \; \chi_k^{\text{A}} = \mathcal{W} - \chi_k^{\text{B}}.
 \end{equation}

 \vspace{-0.3cm}\noindent
It can be seen that  $\chi_k^{\text{A}}$ and $\chi_k^{\text{B}}$ contain non-overlapping elements.
 In Phase II-A,  only $U_k$  is permitted to transmit  pilot symbols in time slots $\xi_k+1\le i \le \xi_k+\gamma$ with $\xi_k = \tau_1+(k-2)\gamma$, and the corresponding  pilot sequence and reflection patterns are set as
  \vspace{-0.3cm}\begin{equation}\label{eq40-1}
  a_{k,i} = \left\{
\begin{array}{ll}
1, &\text{ if } \lceil\frac{i-\tau_1}{\gamma}\rceil+1 = k, \\
0, &\text{ otherwise,}
\end{array}\right. \;\;
{{\theta}}_{n,i} = \left\{\begin{array}{ll}
1, & \text{ if } n \in \mathcal{E}_i \text{ and } i \in [\xi_k+1,\xi_k+\gamma -1], \\
-1, &\text{ if } n \in \chi_k^{\text{A}} \text{ and } i = \xi_k+\gamma, \\
0, & \text{ otherwise,}
\end{array}\right.
 \end{equation}

 \vspace{-0.3cm}\noindent
 where $\mathcal{E}_i \triangleq \big\{ \chi_k^{\text{A}}(\lfloor\frac{i-\xi_k}{\gamma}\rfloor M+1),\cdots,\chi_k^{\text{A}}(\lfloor\frac{i-\xi_k}{\gamma}\rfloor M+M)\big\}$.
{ By defining
\vspace{-0.3cm}
\begin{equation}
{\mathbf{V}}^{\text{II}}_k = \left[\begin{array}{cc}
{\mathbf{H}}_{1}^k\text{diag}(\bar{{\boldsymbol{\theta}}}_{\xi_k+1}) &{{\mathbf{I}}_M} \\
\vdots & \vdots  \\
{\mathbf{H}}_{1}^k\text{diag}(\bar{{\boldsymbol{\theta}}}_{\xi_k+\gamma}) &{{\mathbf{I}}_M}
\end{array}\right],\;
 {\mathbf{H}}_{1}^k = [{\mathbf{h}}_{1,\chi_k^{\text{A}}(1)},\cdots,{\mathbf{h}}_{1,\chi_k^{\text{A}}(N-\delta)}]
,\;\bar{{\boldsymbol{\theta}}}_{i} = [\theta_{\chi_k^{\text{A}}(1),i};\cdots;\theta_{\chi_k^{\text{A}}(N-\delta),i}],
\end{equation}}

\vspace{-1.0cm}
 \noindent the received signal at the BS can be written as
 \vspace{-0.3cm}\begin{equation}
 {\mathbf{y}}^{\text{II}}_k \triangleq [ {\mathbf{y}}^{\xi_k+1};\cdots;{\mathbf{y}}^{\xi_k+\gamma}] =\sqrt{p}{\mathbf{V}}^{\text{II}}_k[{{\mu}}_{k,\chi_k^{\text{A}}(1)},\cdots,{{\mu}}_{k,\chi_k^{\text{A}}(N-\delta)},{\mathbf{h}}_{d,k}^T ]^T,\; \; k \in [2,K].
 \end{equation}

%
\vspace{-0.2cm}\noindent
As a result,   the scaling factors in $\chi_k^{\text{A}}$ and the direct channel associated with $U_k$ can be estimated simultaneously as
 \vspace{-0.3cm}
\begin{equation}\label{eq32-1}
[{\hat{\mu}}_{k,\chi_k^{\text{A}}(1)},\cdots,{\hat{\mu}}_{k,\chi_k^{\text{A}}(N-\delta)},{\hat{\mathbf{h}}}_{d,k}^T ]^T = \frac{1}{\sqrt{p}}({\mathbf{V}}^{\text{II}}_k)^{\dagger}{\mathbf{y}}^{\text{II}}_k.
\end{equation}

\vspace{-0.2cm}
 In Phase II-B,   the pilot sequences of the users and reflection patterns at the IRS in time slot  $i \ge \tau_1+(K-1)\gamma +1$  are designed as
 \vspace{-0.3cm}
 \begin{equation}\label{eq28-1}
a_{k, i}=\left\{\begin{array}{ll}
1, & \text { if } k \in \mathcal{R}_{i},  \\
0, & \text { otherwise},
\end{array}\right. \; \;
\theta_{n, i}=\left\{\begin{array}{ll}
1, & \text { if } n \in \mathcal{W}_{i} \\
0, & \text { otherwise},
\end{array}\right.
 \end{equation}

 \vspace{-0.2cm}\noindent
where
$\mathcal{R}_{i}\triangleq\{\lceil\frac{(i-\tau_1 - (K-1)\gamma-1)M+1}{\delta}\rceil+1,\cdots, \lceil\frac{(i-\tau_1 - (K-1)\gamma-1)M+M_i}{\delta}\rceil+1 \}$,
$\mathcal{W}_{i}\triangleq\{  ((i-\tau_1 - (K-1)\gamma-1)M +m )\%  N, \forall m \in [1,M_i] \}$ and
$M_i = \min(M, (K-1)\delta-(i-\tau_1-(K-1)\gamma-1)M)$.
Then, the received signal can be written as
 \vspace{-0.3cm}
 \begin{equation}\label{B-2}
{\mathbf{y}}^{\left(i\right)}= \sqrt{p}\sum\limits_{k \in \mathcal{R}_{i}}\mathbf{h}_{d,k}+\sqrt{p}\sum\limits_{k \in \mathcal{R}_{i}} \sum\limits_{n \in \mathcal{W}_{i}} \mu_{k, n} \mathbf{h}_{1, n}, \; \;  i \in [\tau_1+(K-1)\gamma+1, \tau],
 \end{equation}

 \vspace{-0.2cm}\noindent
{which includes $|\mathcal{R}_{i}||\mathcal{W}_{i}|$ scaling factors, i.e.,  $\mu_{k,n}, k\in \mathcal{R}_{i}, n\in \mathcal{W}_{i}$. Note that with the designed pilot sequences and reflection patterns in \eqref{eq40-1}, the scaling factors $\mu_{\mathcal{R}_{i}(j_1),\mathcal{W}_{i}(j_2)},j_1,j_2\in [1,M_i], j_1\neq j_2$ and the direct channels $\mathbf{h}_{d,k},k\in \mathcal{R}_{i}$ have been estimated in Phase II-A according to \eqref{eq32-1}, hence
 the rest scaling factors $\mu_{\mathcal{R}_{i}(j),\mathcal{W}_{i}(j)},j\in [1,M_i]$ can be easily estimated as follows:}
 \vspace{-0.6cm}
 \begin{equation}\label{eq30}
[{\hat{\mu}}_{ \mathcal{R}_{i}(1), \mathcal{W}_{i}(1)}; \cdots; {\hat{\mu}}_{\mathcal{R}_{i}(M_i), \mathcal{W}_{i}(M_i)}] = [{\hat{\mathbf{h}}}_{1, \mathcal{W}_{i}(1)}, \cdots, {\hat{\mathbf{h}}}_{1, \mathcal{W}_{i}(M_i)}]^{\dagger} \frac{{\bar{\mathbf{y}}}^{(i)}}{\sqrt{p}},\; \;  i \in [\tau_1+(K-1)\gamma+1, \tau],
 \end{equation}

 \vspace{-0.4cm}\noindent
where
${\bar{\mathbf{y}}}^{(i)}={\mathbf{y}}^{\left(i\right)}-\sum_{k \in \mathcal{R}_{i}}\sqrt{p}{\hat{\mathbf{h}}}_{d,k}-\sum_{k \in \mathcal{R}_{i}} \sum_{n \in \mathcal{W}_{i} \cap \chi_k^{\text{A}}} \sqrt{p} \hat{\mu}_{k, n} {\hat{\mathbf{h}}}_{1, n}$.

{To sum up, according to \eqref{B-1} and \eqref{eq32-1}, the direct channel and  $\min(N,(\gamma -1)M)$ scaling factors associated with $U_k, k\neq 1$, are estimated simultaneously, hence the estimation accuracy of
 the scaling factors  will not be deteriorated by the estimation error of the corresponding direct channel. }

\vspace{-0.4cm}\subsection{Overall Channel Training Overhead}
According to the above two subsections, the overall channel training overhead required by the proposed 2PCE strategy is
 \vspace{-0.3cm}
 \begin{equation}
\tau = \tau_1 + \tau_2 = \left\{\begin{array}{ll}
N+2K-1, &\text{ if } M\ge N, \\
N+K+\lceil \frac{(K-1)N}{M} \rceil, &\text{ otherwise, }
\end{array}\right.
 \end{equation}

 \vspace{-0.3cm}\noindent
which is the same with that of the 3PCE strategy in \cite{arXiv1912}, i.e, $ K+N+\max(K-1,\lceil\frac{(K-1) N}{M}\rceil)$.
Note that both the proposed 2PCE strategy and the 3PCE strategy in \cite{arXiv1912} are feasible solutions for achieving the minimum required channel training overhead in the considered IRS-aided multiuser SIMO system, the difference is that the proposed 2PCE strategy contains less estimation phases and thus the error propagation can be well-controlled.

\vspace{-0.4cm}\section{ Asymptotic MSE analysis}\label{MSEsection}
In the previous section, we have shown how to perfectly estimate all the channels in detail for the ideal case without receive noise at the BS.
In this section, we consider the practical case with noise and present an LS estimator for the proposed 2PCE strategy.\footnote{Note that when the estimation error of the reflected channel associated with the typical user is not neglected, the derivation of  the linear minimum MSE (LMMSE) estimator can be very complex, hence for simplicity we only consider the LS estimation in this work. Further investigation into  more complicated estimators are left for future work. }  Moreover, we derive  the  asymptotic MSE of the estimated $\mathbf{H}_1$, $\{\mathbf{h}_{d,k} \}_{k=1}^K$ and $\{\bm{\mu}_k\}_{k=2}^K$ when $M$ becomes large to investigate how  the error propagation issue is alleviated by the proposed 2PCE strategy.
 Besides, performance comparison between the proposed 2PCE strategy and the 3PCE strategy  \cite{arXiv1912} in terms of the asymptotic MSE is provided to show the advantages of the proposed strategy. In the sequel,
 for ease of analysis, we assume that both the IRS-BS and user-IRS channels  follow Rayleigh
distribution, i.e., $\mathbf{g}_n\sim \mathcal{CN}(\mathbf{0},l^{\textrm{IB}}\mathbf{I})$ and $\mathbf{h}_{r,k}\sim \mathcal{CN}(\mathbf{0}, l^{\textrm{UI}}_k\mathbf{I})$, where $l^{\textrm{IB}}$ ($l^{\textrm{IU}}_k$) represents the path loss of the link from  the IRS to the BS  (from $U_k$ to the IRS).
In addition, some necessary  propositions and lemmas used throughout this section are listed in Appendix A. 
\vspace{-0.4cm}\subsection{MSE Analysis in Phase I}\label{V-A}
With receive noise at the BS, the received signal at the BS given in \eqref{eq44} can be rewritten as
 \vspace{-0.3cm}
 \begin{equation}\label{eq49}
{\mathbf{Y}}^{\text{I}} = \sqrt{p}{\mathbf{X}}^{\text{I}}{\mathbf{V}}^{\text{I}}+{\mathbf{N}}^{\text{I}},
 \end{equation}

 \vspace{-0.3cm}\noindent
where ${\mathbf{N}}^{\text{I}} = [{\mathbf{n}}^{(1)},\cdots,{\mathbf{n}}^{(N+1)}]\in {\mathbb{C}}^{M\times (N+1)}$.
By right-multiplying $\mathbf{Y}^{\text{I}}$ by the inverse of ${\mathbf{V}}^{\text{I}}$ (or equivalently, $\frac{(\mathbf{V}^{\text{I}})^H}{N+1}$), we can obtain the LS estimation of $\mathbf{h}_{d,1}$  and $\mathbf{H}_1$  as 
 \vspace{-0.3cm}
 \begin{equation}\label{eq86}
\hat{{\mathbf{X}}^{\text{I}}} \triangleq [\hat{\mathbf{h}}_{d,1},\hat{\mathbf{H}}_1] = {\mathbf{X}}^{\text{I}} + \frac{{\mathbf{N}}^{\text{I}}({\mathbf{V}}^{\text{I}})^{H}}{\sqrt{p}(N+1)}.
 \end{equation}

 \vspace{-0.2cm}\noindent
From \eqref{eq86}, it can be observed that the estimation errors of the direct and reflected channels associated with $U_1$
  have complex Gaussian entries since the channel noise is modeled as AWGN
with i.i.d. entries and the adopted LS estimator only involves linear operations to $\mathbf{Y}^{\text{I}}$. Besides, we can see that the CSI error matrix can be expressed as
 \vspace{-0.3cm}
 \begin{equation}\label{A-8}
 \Delta {\mathbf{X}}^{\text{I}} \triangleq [\Delta \mathbf{h}_{d,1},\Delta \mathbf{H}_1]  = {\mathbf{X}}^{\text{I}} - \hat{{\mathbf{X}}^{\text{I}}}  =\frac{{\mathbf{N}}^{\text{I}}({\mathbf{V}}^{\text{I}})^{H}}{\sqrt{p}(N+1)},
 \end{equation}

 \vspace{-0.3cm}\noindent
 which satisfies $\mathbb{E}\{\Delta {\mathbf{X}}^{\text{I}}\} = {\mathbf{0}}$ and
$
 \mathbb{E}\{\Delta {\mathbf{X}}^{\text{I}}(\Delta {\mathbf{X}}^{\text{I}})^H\} = \frac{\sigma^2}{p}\mathbf{I}_M
$.
As a result, the MSEs for estimating  $\mathbf{h}_{d,1}$ and $\mathbf{H}_1$  are respectively given by
 \vspace{-0.3cm}
 \begin{equation}\label{E21}
\epsilon_{d,1,\text{2P}} \triangleq \mathbb{E}\big\{\text{tr}(\Delta {\mathbf{h}}_{d,1}\Delta {\mathbf{h}}_{d,1}^H)\big\} = \frac{M\sigma^2}{p(N+1)},\;\;
\epsilon_{r,1,\text{2P}} \triangleq \mathbb{E}\big\{\text{tr}(\Delta {\mathbf{H}}_{1}\Delta {\mathbf{H}}_{1}^H)\big\} = \frac{MN\sigma^2}{p(N+1)}.
\end{equation}

 \vspace{-0.2cm}\noindent
{From \eqref{E21}, we can see that $\epsilon_{d,1,\text{2P}}$ decreases with the increasing of $N$ since the estimation of $\mathbf{h}_{d,1}$ is obtained based on the received signals in a total number of $N+1$ time slots, hence the MSE for estimating $\mathbf{h}_{d,1}$ in the 2PCE strategy is expected to be lower than that in the 3PCE strategy which only uses one time slot.} Besides, we can observe that $\epsilon_{d,1,\text{2P}}$ and $\epsilon_{r,1,\text{2P}}$ are not related, which means that the estimation of the reflected  channel associated with the typical user will not be affected by that of the corresponding direct channel.
\vspace{-0.5cm}\subsection{Asymptotic MSE Analysis in Phase II}\label{V-B}
In Phase II, the direct and reflected channels associated with the non-typical users are estimated simultaneously to alleviate the negative
effects caused by error propagation.
 By replacing $\mathbf{H}_1$ with $\hat{\mathbf{H}}_1 + \Delta \mathbf{H}_1$, the effective received signal at the BS given in \eqref{eq36} can be re-expressed as
 \vspace{-0.3cm}
 \begin{equation}\label{eq46}
 {\mathbf{y}}^{(i)}=\sqrt{p}\sum\limits_{k=2}^K a_{k,i}({\mathbf{h}}_{d,k}+{\hat{\mathbf{H}}}_{1}\text{diag}({\boldsymbol{\theta}}_i){\bm{\mu}}_k) + \sqrt{p}\sum\limits_{k=2}^K a_{k,i}\Delta {\mathbf{H}}_{1}\text{diag}({\boldsymbol{\theta}}_i){\bm{\mu}}_k + {\mathbf{n}}^{(i)}, \;\;
 i \in [\tau_1+1,\tau_1+\tau_2].
 \end{equation}

 \vspace{-0.3cm}\noindent
 Then, the overall received signal at the BS with receive noise can be written as
 \vspace{-0.3cm}
 \begin{equation}\label{eq53}
{\mathbf{y}}^{\text{II}}\triangleq[{\mathbf{y}}^{(\tau_1)};\cdots; {\mathbf{y}}^{(\tau_1+\tau_2)}] = \sqrt{p}{\hat{\mathbf{W}}}^{\text{II}}{\mathbf{x}}^{\text{II}} + \sqrt{p}{\mathbf{W}}_{\Delta {\mathbf{H}}_{1}}[{\bm{\mu}}_2;\cdots;{\bm{\mu}}_K]+{\mathbf{n}}^{\text{II}},
 \end{equation}

 \vspace{-0.2cm}\noindent
where ${\mathbf{n}}^{\text{II}} = [{\mathbf{n}}^{(\tau_1+1)};\cdots;{\mathbf{n}}^{(\tau_1+\tau_2)}]$,
 \vspace{-0.3cm}
 \begin{equation}\label{eq48}
{\hat{\mathbf{W}}}^{\text{II}} = \left[\begin{array}{cccccc}
a_{2,\tau_1+1}{\hat{\mathbf{H}}}_{1}\text{diag}({\boldsymbol{\theta}}_{\tau_1+1})&\cdots &a_{K,\tau_1+1}{\hat{\mathbf{H}}}_{1}\text{diag}({\boldsymbol{\theta}}_{\tau_1+1}) &a_{2,\tau_1+1}{{\mathbf{I}}_M} &\cdots &a_{K,\tau_1+1}{{\mathbf{I}}_M}\\
\vdots & \ddots & \vdots & \vdots & \ddots & \vdots \\
a_{2,\tau_1+\tau_2}{\hat{\mathbf{H}}}_{1}\text{diag}({\boldsymbol{\theta}}_{\tau_1+\tau_2})&\cdots &a_{K,\tau_1+\tau_2}{\hat{\mathbf{H}}}_{1}\text{diag}({\boldsymbol{\theta}}_{\tau_1+\tau_2}) &a_{2,\tau_1+\tau_2}{{\mathbf{I}}_M} &\cdots &a_{K,\tau_1+\tau_2}{{\mathbf{I}}_M}
\end{array}\right],
\end{equation}
 \vspace{-0.8cm}\begin{equation}\label{eq47}
{\mathbf{W}}_{\Delta {\mathbf{H}}_{1}} = \left[\begin{array}{ccc}
a_{2,\tau_1+1}\Delta{\mathbf{H}}_{1}\text{diag}({\boldsymbol{\theta}}_{\tau_1+1})&\cdots &a_{K,\tau_1+1}\Delta{\mathbf{H}}_{1}\text{diag}({\boldsymbol{\theta}}_{\tau_1+1}) \\
\vdots & \ddots & \vdots  \\
a_{2,\tau_1+\tau_2}\Delta{\mathbf{H}}_{1}\text{diag}({\boldsymbol{\theta}}_{\tau_1+\tau_2})&\cdots &a_{K,\tau_1+\tau_2}\Delta{\mathbf{H}}_{1}\text{diag}({\boldsymbol{\theta}}_{\tau_1+\tau_2})
\end{array}\right].
 \end{equation}

 \vspace{-0.2cm}\noindent
Similar to Section \ref{III-B}, the asymptotic MSE analysis of the proposed 2PCE strategy in Phase II are also divided in  two cases: $M \ge N$ and  $M \textless N$.
\subsubsection{$M\ge N$}
 In order to estimate the scaling vector $\bm{\mu}_k$ and the direct channel ${\mathbf{h}}_{d,k}$, we left-multiply ${\mathbf{y}}^{\text{II}}$  with the auxiliary matrix ${\bm{\Sigma}}_k$ (given in \eqref{SIGMA})  and thereby obtain the following equation:
 \vspace{-0.3cm}
 \begin{equation}
 {\bm{\Sigma}}_k {\mathbf{y}}^{\text{II}}=\sqrt{p}(K-1)({\hat{\mathbf{V}}}^{\text{II}} [{\bm{\mu}}_k;
 {\mathbf{h}}_{d,k}] + \Delta{\mathbf{S}}{\bm{\mu}}_k) + {\bm{\Sigma}}_k{\mathbf{n}}^{\text{II}},\;\; k\in [2,K],
\end{equation}

 \vspace{-0.2cm}\noindent where ${\hat{\mathbf{V}}}^{\text{II}}  = [{\hat{\mathbf{H}}}_1, {{\mathbf{I}}_M}; -{\hat{\mathbf{H}}}_1,{{\mathbf{I}}_M}] $
 and
 $\Delta{{\mathbf{S}}} =[
\Delta{\mathbf{H}}_1;
 -\Delta{\mathbf{H}}_1
 ]$.
Therefore, the LS estimation of ${\bm{\mu}}_k$ and ${\mathbf{h}}_{d,k}$ can be obtained as follows:
 \vspace{-0.4cm}
 \begin{equation}\label{A-16}
[{\hat{\bm{\mu}}}_k;
 {\hat{\mathbf{h}}}_{d,k}]
  = [{\bm{\mu}}_k;
 {\mathbf{h}}_{d,k}] +  \underbrace{ \frac{1}{\sqrt{p}(K-1)}({\hat{\mathbf{V}}}^{\text{II}})^{\dagger}{\bm{\Sigma}}_k{\mathbf{n}}^{\text{II}}+({\hat{\mathbf{V}}}^{\text{II}})^{\dagger} \Delta{\mathbf{S}}{\bm{\mu}}_k }_{\Delta \mathbf{x}_k^{\text{II}}},\;\; k \in [2,K],
 \end{equation}

 \vspace{-0.4cm}\noindent
where $\Delta \mathbf{x}_k^{\text{II}}$ denotes the composite estimation error matrix in Phase II. It can be seen from \eqref{A-16} that the estimation accuracy in this phase is
closely related to two factors, i.e., the channel noise $\mathbf{n}^{\text{II}}$ and the estimation error of $\mathbf{H}_1$.
 Since each entry of $\mathbf{n}^{\text{II}}$ and $\Delta \mathbf{S}$ follows independent  complex normal distribution, the covariance matrix of $\Delta {\mathbf{x}}^{\text{II}}_k$ can be decomposed into two parts, i.e.,
 \vspace{-0.4cm}
 \begin{equation}\label{eq54}
\begin{array}{l}
\mathbb{E}\big\{\Delta {\mathbf{x}}^{\text{II}}_k(\Delta {\mathbf{x}}^{\text{II}}_k)^H\big\}
= \mathbf{P}_{k,1}^{\text{a}}  + \mathbf{P}_{k,2}^{\text{a}},
\end{array}
 \end{equation}

 \vspace{-0.2cm}\noindent
where  $\mathbf{P}_{k,1}^{\text{a}}  \triangleq \frac{\mathbb{E}\{ ({\hat{\mathbf{V}}}^{\text{II}})^{\dagger}{\bm{\Sigma}}_k{\mathbf{n}}^{\text{II}}({\mathbf{n}}^{\text{II}}{\bm{\Sigma}}_k)^H (({\hat{\mathbf{V}}}^{\text{II}})^H)^{\dagger}\}}{p(K-1)^2}$  and $\mathbf{P}_{k,2}^{\text{a}} \triangleq \mathbb{E}\big\{({\hat{\mathbf{V}}}^{\text{II}})^{\dagger} \Delta{\mathbf{S}}{\bm{\mu}}_k {\bm{\mu}}_k^H \Delta{\mathbf{S}}^H (({\hat{\mathbf{V}}}^{\text{II}})^{\dagger})^H \big\}$.
{Note that the first part $\mathbf{P}_{k,1}^{\text{a}}$ in \eqref{eq54} is  related to  both $\mathbf{n}^{\text{II}}$ and $\Delta \mathbf{H}_1$ (contained in ${\hat{\mathbf{V}}}^{\text{II}}$), while the second part $\mathbf{P}_{k,2}^{\text{a}}$ is only related to $\Delta \mathbf{H}_1$ (contained in ${\hat{\mathbf{V}}}^{\text{II}}$ and $\Delta \mathbf{S}$).
 The exact characterization of $\mathbf{P}_{k,1}^{\text{a}}$ and $\mathbf{P}_{k,2}^{\text{a}}$ is very challenging since the pseudo-inverse operation of the random matrix ${\hat{\mathbf{V}}}^{\text{II}}$ or $({\hat{\mathbf{V}}}^{\text{II}})^{H}$ is involved.\footnote{Note that in [23], the MSE performance of the LMMSE estimator is derived under the assumption that the reflected channel associated with the typical user is perfectly known. In this work, we focus on the more practical case, where no such assumption is made. }
In the following, we derive trackable forms for these two parts based on  random matrix theory and some necessary lemmas presented in  Appendix A.}

First, we focus on the derivation of
 $\mathbf{P}_{k,1}^{\text{a}}$. Since the entries of $\mathbf{n}^{\text{II}}$ and $(\hat{\mathbf{V}}_{\text{II}})^{\dagger}$ are  independent with each other, $\mathbf{P}_{k,1}^{\text{a}}$ can be equivalently transformed into $\mathbf{P}_{k,1}^{\text{a}}
=\frac{ \sigma^2}{p(K-1)}\mathbb{E}\big\{\big(({\hat{\mathbf{V}}}^{\text{II}})^H {\hat{\mathbf{V}}}^{\text{II}}\big)^{-1}\big\}$. Moreover, we observe that
$({\hat{\mathbf{V}}}^{\text{II}})^H {\hat{\mathbf{V}}}^{\text{II}} =
2[
\hat{{\mathbf{H}}}_1^H{\hat{\mathbf{H}}}_1,\mathbf{0}_{N \times M};
\mathbf{0}_{M \times N},{\mathbf{I}}_M
]$,
which  is a $2\times 2$ block matrix,   therefore the inverse of $({\hat{\mathbf{V}}}^{\text{II}})^H {\hat{\mathbf{V}}}^{\text{II}}$ can be obtained according to Lemma \ref{lemma1} and $\mathbf{P}_{k,1}^{\text{a}}$ can be further simplified  as follows:
 \vspace{-0.3cm}
 \begin{equation}\label{D-3}
\begin{aligned}
\mathbf{P}_{k,1}^{\text{a}}
= \frac{ \sigma^2}{2p(K-1)} \left[\begin{array}{cc}  \mathbb{E}\big\{(\hat{{\mathbf{H}}}_1^H{\hat{\mathbf{H}}}_1)^{-1}\big\} & {\mathbf{0}}_{N\times M}\\
{\mathbf{0}}_{M\times N} & {\mathbf{I}}_M
\end{array} \right].
\end{aligned}
 \end{equation}

 \vspace{-0.3cm}\noindent
As for $\mathbf{P}_2^{\text{a}}$, we first define
$\mathbf{M}_1 \triangleq \Delta{\mathbf{S}}{\bm{\mu}}_k {\bm{\mu}}_k^H \Delta{\mathbf{S}}^H = [
  \mathbf{E}_{k}, -\mathbf{E}_{k};
  -\mathbf{E}_{k},\mathbf{E}_{k}]$
  where $\mathbf{E}_{k} = \Delta\mathbf{H}_1 \bm{\mu}_k\bm{\mu}_k^H$ $\Delta\mathbf{H}_1^H$, then the pseudo-inverse of $\mathbf{M}_1$ can be obtained by
  $\frac{1}{4} [
     \mathbf{E}_{k}^{\dagger}, -\mathbf{E}_{k}^{\dagger};
  -\mathbf{E}_{k}^{\dagger}, \mathbf{E}_{k}^{\dagger}
   ]$
  according to Lemma \ref{lemma2}. As a result,  $\mathbf{P}_{k,2}^{\text{a}}$ can be simplified as
  \vspace{-0.4cm}
  \begin{equation}\label{D-2}
\begin{aligned}
\mathbf{P}_{k,2}^{\text{a}}
 = \mathbb{E}\big\{\big(({\hat{\mathbf{V}}}^{\text{II}})^H
  \mathbf{M}_1^{\dagger}{\hat{\mathbf{V}}}^{\text{II}}
  \big)^{\dagger}\big\}
    = \left[ \begin{array}{cc}\mathbb{E}\{ \mathbf{R}_k  \} & \mathbf{0}_{N\times M}\\
   \mathbf{0}_{M \times N} &  \mathbf{0}_{M \times M}  \end{array}\right],
  \end{aligned}
 \end{equation}

 \vspace{-0.3cm}\noindent
where $\mathbf{R}_k = \hat{\mathbf{H}}_1^{\dagger}\Delta\mathbf{H}_1 \bm{\mu}_k\bm{\mu}_k^H\Delta\mathbf{H}_1^H(\hat{\mathbf{H}}_1^H)^{\dagger}$.
Based on \eqref{D-3} and \eqref{D-2}, the covariance matrix of $\Delta {\mathbf{x}}^{\text{II}}_k$ in \eqref{eq54} can be rewritten as
 \vspace{-0.3cm}
 \begin{equation}
\mathbb{E}\big\{\Delta {\mathbf{x}}^{\text{II}}_k(\Delta {\mathbf{x}}^{\text{II}}_k)^H\big\}
= \frac{\sigma^2}{2p(K-1)} \left[\begin{array}{cc}  \mathbb{E}\big\{(\hat{{\mathbf{H}}}_1^H{\hat{\mathbf{H}}}_1)^{-1}\big\} & {\mathbf{0}}_{N\times M}\\
{\mathbf{0}}_{M\times N} & {\mathbf{I}}_M
\end{array} \right]+ \left[ \begin{array}{cc}\mathbb{E}\{ \mathbf{R}_k \} & \mathbf{0}_{N\times M}\\
   \mathbf{0}_{M \times N} &  \mathbf{0}_{M \times M}  \end{array}\right].
 \end{equation}

 \vspace{-0.2cm}\noindent
Accordingly, the MSEs for estimating  $\mathbf{h}_{d,k}$ and $\bm{\mu}_k, k \in [2,K]$, are given by
\vspace{-0.2cm}\begin{subequations}
\begin{equation}\label{E2k}
\epsilon_{d,k,\text{2P}}^{\text{a}} \triangleq \mathbb{E} \big\{\text{tr}( \Delta {\mathbf{h}}_{d,k} \Delta {\mathbf{h}}_{d,k}^H )\big\}
 =  \frac{M\sigma^2}{2p(K-1)},
\end{equation}
\vspace{-0.8cm}
\begin{equation}\label{D-5}
\begin{aligned}
& \epsilon_{\bm{\mu},k,\text{2P}}^{\text{a}} \triangleq \mathbb{E}\big\{\text{tr}( \Delta \bm{\mu}_{k} \Delta \bm{\mu}_k^H)\big\} =\frac{\sigma^2}{2p(K-1)}\mathbb{E}\big\{ \text{tr}\big((\hat{{\mathbf{H}}}_1^H{\hat{\mathbf{H}}}_1)^{-1}\big)\big\} + \mathbb{E}\big\{\text{tr}(\mathbf{R}_k)\big\}.
\end{aligned}
\end{equation}
\end{subequations}

\vspace{-0.2cm}\noindent
{It can be seen from \eqref{E2k} that $\epsilon_{d,k,\text{2P}}^{\text{a}}$  is only affected by  the channel noise in Phase II and it is
inversely proportional to $2(K-1)$
  since the estimation of $\mathbf{h}_{d,k}$ is obtained based on the received signals in a total number of $2(K-1)$ time slots.} Besides,  $\epsilon_{\bm{\mu},k,\text{2P}}^{\text{a}}$ is further related to the CSI error matrix of $\mathbf{H}_1$,
which is still difficult to characterize since it involves the inverse operation of $\hat{{\mathbf{H}}}_1^H{\hat{\mathbf{H}}}_1$.

Let $\mathbf{P}_{\bm{\mu}_k,1}^{\text{a}} \triangleq \frac{\sigma^2}{2p(K-1)}\mathbb{E}\big\{ \text{tr}\big((\hat{{\mathbf{H}}}_1^H{\hat{\mathbf{H}}}_1)^{-1}\big)\big\}$ and $\mathbf{P}_{\bm{\mu}_k,2}^{\text{a}} \triangleq \mathbb{E}\big\{\text{tr}(\mathbf{R}_k)\big\}$ denote the two terms on the RHS of \eqref{D-5}, in the following, we focus on further simplifying $\mathbf{P}_{\bm{\mu}_k,1}^{\text{a}}$ and $\mathbf{P}_{\bm{\mu}_k,2}^{\text{a}}$  when $M$ becomes large to draw useful insights and we have the following proposition.
\vspace{-0.3cm}
\begin{proposition}\label{Pro2}\em
{ $\mathbf{P}_{\bm{\mu}_k,1}^{\text{a}}$ and $\mathbf{P}_{\bm{\mu}_k,2}^{\text{a}}$ can be approximated by
$
 \mathbf{P}_{\bm{\mu}_k,1}^{\text{a}} \approx \frac{\sigma^2(N+1)N}{2(K-1)M\big(p(N+1) l^{\text{UI}}_1l^{\text{IB}} + \sigma^2\big)}$ and $\mathbf{P}_{\bm{\mu}_k,2}^{\text{a}}
 \approx  \frac{M\sigma^2\mathbb{E}\{\bm{\mu}_k^H \bm{\mu}_k\}}{Nl^{\text{UI}}_1l^{\text{IB}}p(N+1)+M\sigma^2}
$ when $M$ becomes asymptotically large.}
\end{proposition}
\vspace{-0.3cm}
\begin{proof}
Please refer to Appendix B. 
\end{proof}
According to Proposition \ref{Pro2}, by substituting the asymptotic values of $ \mathbf{P}_{\bm{\mu}_k,1}^{\text{a}}$ and $ \mathbf{P}_{\bm{\mu}_k,2}^{\text{a}}$ into \eqref{D-5}, we can obtain the asymptotic MSE for estimating  $\bm{\mu}_k$ as follows:
 \vspace{-0.3cm}\begin{equation}\label{D-7}
\begin{aligned}
& \epsilon_{\bm{\mu}_k,\text{2P}}^{\text{a}} \approx
\frac{\sigma^2(N+1)N}{2(K-1)M\big(p(N+1) l^{\text{UI}}_1l^{\text{IB}} + \sigma^2\big)} + \frac{M\sigma^2\mathbb{E}\{\bm{\mu}_k^H \bm{\mu}_k\}}{Nl^{\text{UI}}_1l^{\text{IB}}p(N+1)+M\sigma^2}.
\end{aligned}
 \end{equation}

 \vspace{-0.2cm}\noindent
{From \eqref{D-7}, we can observe that $\epsilon_{\bm{\mu}_k,\text{2P}}^{\text{a}}$ is  inversely proportional to the transmit power $p$, the path losses of the links from the typical user to the IRS and from  the IRS to the BS, i.e.,  $l^{\text{UI}}_1$ and $l^{\text{IB}}$, while it is directly proportional to  $\mathbb{E}\{\bm{\mu}_k^H \bm{\mu}_k\}$. The first term on the RHS of \eqref{D-7} will dominate the second term when $N$ becomes asymptotically large, and $\epsilon_{\bm{\mu}_k,\text{2P}}^{\text{a}}$ is directly proportional to $N$ in this case.  Besides, since the first term on the RHS of  \eqref{D-7} is affected by the channel noise contained in the $2(K-1)$ received signals, $ \epsilon_{\bm{\mu}_k,\text{2P}}^{\text{a}}$ decreases with the increasing  of $K$.}
\subsubsection{$M\textless N$}
In this case, we assume for simplicity an orthogonal channel training strategy, i.e., the users do not share the time slots and only one user is allowed to transmit its pilot symbol to the BS in each time slot.\footnote{
 Note that in the general case when the users are allowed to share some of the time slots, the asymptotic MSE analysis can be very complicated. In this paper, we focus on this simplified case, and further investigation into the more general case is left for future work.     }
 Specifically, in time slot $i\in [\xi'_k + 1,\xi'_k+\gamma + 1]$ with $\xi'_k = \tau_1 + (\gamma+1)(k-1)$, only $U_k$ is allowed to transmit pilot symbol $a_{k,i}=1$ and some  selected reflecting elements  are switched on, then the direct channel and corresponding scaling factors associated with $U_k$ can be estimated. The designed reflection patterns in time slot $i \in [\xi'_k + 1,\xi'_k+\gamma + 1]$ are set as
 \vspace{-0.3cm}\begin{equation}
 \theta_{n,i} = \left\{
 \begin{array}{ll}
 1, & \text{ if } i - \xi'_k = \lceil\frac{n}{M}\rceil, \\
 -1, & \text{ if } i - \xi'_k = \gamma +1,\\
 0, & \text{ otherwise. }
 \end{array}
 \right.
 \end{equation}

 \vspace{-0.2cm}\noindent
Let us define
$\Delta{{\mathbf{S}}}_k = [\Delta{{\mathbf{H}}}_{1}\text{diag}({{\boldsymbol{\theta}}}_{\xi'_k+1});\cdots; \Delta{{\mathbf{H}}}_{1}\text{diag}({{\boldsymbol{\theta}}}_{\xi'_k+\gamma+1})  ] $,
\vspace{-0.3cm}$$
{\mathbf{Q}}^{\text{II}}_k
= \left[\begin{array}{cc}
{{\mathbf{H}}}_{1}\text{diag}({{\boldsymbol{\theta}}}_{\xi'_k+1}) &{{\mathbf{I}}_M} \\
\vdots & \vdots  \\
{{\mathbf{H}}}_{1}\text{diag}({{\boldsymbol{\theta}}}_{\xi'_k+\gamma+1}) &{{\mathbf{I}}_M}
\end{array}\right] \; \text{and} \;
{\hat{\mathbf{Q}}}^{\text{II}}_k= \left[\begin{array}{cc}
{\hat{\mathbf{H}}}_{1}\text{diag}({{\boldsymbol{\theta}}}_{\xi'_k+1}) &{{\mathbf{I}}_M} \\
\vdots & \vdots  \\
{\hat{\mathbf{H}}}_{1}\text{diag}({{\boldsymbol{\theta}}}_{\xi'_k+\gamma+1}) &{{\mathbf{I}}_M}
\end{array}\right],
$$

\vspace{-0.3cm}\noindent
then the overall received signals in time slots $i\in [\xi'_k + 1,\xi'_k+\gamma + 1]$ can be expressed as
 \vspace{-0.35cm}\begin{equation}\label{A-15}
  {\mathbf{y}}^{\text{II}}_k\triangleq [\mathbf{y}^{(\xi'_k+1)};\cdots;\mathbf{y}^{(\xi'_k+\gamma + 1)}] = \sqrt{p}{\mathbf{Q}}^{\text{II}}_k[{\bm{\mu}}_k;{\mathbf{h}}_{d,k}]+ {\mathbf{n}}_k^{\text{II}},\; \; k \in [2,K],
 \end{equation}

 \vspace{-0.3cm}\noindent
where ${\mathbf{n}}_k^{\text{II}} = [{\mathbf{n}}^{(\xi'_k+1)};\cdots;{\mathbf{n}}^{(\xi'_k+\gamma+1)}]$.
 According to \eqref{A-15}, the LS estimation of ${\bm{\mu}}_k$ and ${{\mathbf{h}}}_{d,k}$ associated with $U_k$  are given by
 \vspace{-0.35cm}\begin{equation}\label{A-17}
[\hat{{\bm{\mu}}}_k;{\hat{\mathbf{h}}}_{d,k} ] = [{\bm{\mu}}_k;{{\mathbf{h}}}_{d,k} ] +({\hat{\mathbf{Q}}}^{\text{II}}_k)^{\dagger}\Delta{\mathbf{S}}_k {\bm{\mu}}_k +  \frac{1}{\sqrt{p}} ({\hat{\mathbf{Q}}}^{\text{II}}_k)^{\dagger} {\mathbf{n}}_k^{\text{II}}.
 \end{equation}

 \vspace{-0.3cm}\noindent
Let $\Delta {\mathbf{x}}_k \triangleq  \frac{1}{\sqrt{p}} ({\hat{\mathbf{Q}}}^{\text{II}}_k)^{\dagger} {\mathbf{n}}_k^{\text{II}}+ ({\hat{\mathbf{Q}}}^{\text{II}}_k)^{\dagger}\Delta{\mathbf{S}}_k {\bm{\mu}}_k$ denote the CSI error matrix in \eqref{A-17}, where the first term $ \frac{1}{\sqrt{p}} ({\hat{\mathbf{Q}}}^{\text{II}}_k)^{\dagger} {\mathbf{n}}_k^{\text{II}}$ is caused by  the channel  noise and the estimated reflected channel associated with the typical user $\hat{\mathbf{H}}_1$ obtained in Phase I, and the second term $({\hat{\mathbf{Q}}}^{\text{II}}_k)^{\dagger}\Delta{\mathbf{S}}_k {\bm{\mu}}_k$ is only related to $\hat{\mathbf{H}}_1$. Then, similar to \eqref{eq54},  we can obtain the covariance matrix of $\Delta {\mathbf{x}}_k$ as follows: 
 \vspace{-0.4cm}\begin{equation}\label{eq59}
\begin{array}{l}
\mathbb{E}\{\Delta {\mathbf{x}}_k \Delta {\mathbf{x}}_k^H\}
=\mathbf{P}_{k,1}^{\text{b}} + \mathbf{P}_{k,2}^{\text{b}},
\end{array}
 \end{equation}

 \vspace{-0.2cm}\noindent
where $\mathbf{P}_{k,1}^{\text{b}} \triangleq \frac{1}{p}\mathbb{E}\big\{({\hat{\mathbf{Q}}}^{\text{II}}_k)^{\dagger}{\mathbf{n}}^{\text{II}}_k({\mathbf{n}}^{\text{II}}_k)^H \big(({\hat{\mathbf{Q}}}^{\text{II}}_k)^H\big)^{\dagger}\big\}$ and $\mathbf{P}_{k,2}^{\text{b}} \triangleq \mathbb{E}\big\{({\hat{\mathbf{Q}}}^{\text{II}}_k)^{\dagger} \Delta{\mathbf{S}}_k{\bm{\mu}}_k {\bm{\mu}}_k^H  \Delta{\mathbf{S}}_k^H (({\hat{\mathbf{Q}}}^{\text{II}}_k)^{\dagger})^H \big\}$.

 Similar to the case of $M \ge N$, we will simplify $\mathbf{P}_{k,1}^{\text{b}}$ and $\mathbf{P}_{k,2}^{\text{b}}$ in the following.
  First,  since $\mathbb{E}\big\{{\mathbf{n}}^{\text{II}}_k({\mathbf{n}}^{\text{II}}_k)^H\big\}= \sigma^2{\mathbf{I}}$, we obtain
$
 \mathbf{P}_{k,1}^{\text{b}}= \frac{\sigma^2}{p}\mathbb{E}\big\{\big(({\hat{\mathbf{Q}}}^{\text{II}}_k)^H {\hat{\mathbf{Q}}}^{\text{II}}_k\big)^{-1}\big\}
$.
Besides, as  $({\hat{\mathbf{Q}}}^{\text{II}}_k)^H {\hat{\mathbf{Q}}}^{\text{II}}_k$ is a $2\times 2$ block matrix which can be expressed as $({\hat{\mathbf{Q}}}^{\text{II}}_k)^H {\hat{\mathbf{Q}}}^{\text{II}}_k = [
\hat{\mathbf{S}}_k^H\hat{\mathbf{S}}_k, \mathbf{0}_{N\times M};
\mathbf{0}_{M\times N}, (\gamma +1) \mathbf{I}_M
]$
%
 with
$
\hat{\mathbf{S}}_k = [ {\hat{\mathbf{H}}}_{1}\text{diag}({{\boldsymbol{\theta}}}_{\xi_k+1});
\cdots;{\hat{\mathbf{H}}}_{1}\text{diag}({{\boldsymbol{\theta}}}_{\xi_k+\gamma+1})]
$,
we can further transform $\mathbf{P}_{k,1}^{\text{b}}$ into
 \vspace{-0.3cm}\begin{equation}\label{A-18}
\mathbf{P}_{k,1}^{\text{b}}  {=}\frac{\sigma^2}{p} \left[
\begin{array}{cc}
\mathbb{E}\big\{( \hat{\mathbf{S}}_k^H\hat{\mathbf{S}}_k)^{-1}\big\}  & \mathbf{0}_{N\times M}  \\
\mathbf{0}_{M\times N}  & \frac{1}{\gamma +1} \mathbf{I}_M
\end{array}\right],
 \end{equation}

 \vspace{-0.2cm}\noindent
according to  Lemma \ref{lemma1}.
Moreover, by employing the result in Lemma \ref{lemma2}, $\mathbf{P}_{k,2}^{\text{b}}$ can be re-expressed as follows:
 \vspace{-0.35cm}\begin{equation}\label{eq82}
\begin{aligned}
&\mathbf{P}_{k,2}^{\text{b}}
= \mathbb{E}\big\{ \big( ({\hat{\mathbf{Q}}}^{\text{II}}_k)^H  (\Delta{\mathbf{S}_k}{\bm{\mu}}_k {\bm{\mu}}_k^H  \Delta{\mathbf{S}}^H_k)^{\dagger} {\hat{\mathbf{Q}}}^{\text{II}}_k  \big)^{\dagger}  \big\}
 =
\left[ \begin{array}{cc}
\mathbb{E}\big\{{\hat{\mathbf{S}}}_k^{\dagger} \Delta{\mathbf{S}}_k{\bm{\mu}}_k {\bm{\mu}}_k^H \Delta{\mathbf{S}}_k^H ({\hat{\mathbf{S}}}_k^{\dagger})^H\big\}& \mathbf{0}_{N\times M}\\
\mathbf{0}_{M\times N}  & \mathbf{0}_{M\times M}
\end{array}  \right].
  \end{aligned}
 \end{equation}

 \vspace{-0.3cm}\noindent
Based on the simplified $\mathbf{P}_{k,1}^{\text{b}}$ and $\mathbf{P}_{k,2}^{\text{b}}$ in \eqref{A-18} and \eqref{eq82},
the  MSEs for estimating $\mathbf{h}_{d,k}$ and $\bm{\mu}_k, k\in[2,K]$, are given  by
 \vspace{-0.4cm}\begin{subequations}\label{A-19}
\begin{equation}\label{L6-13}
\epsilon_{d,k,\text{2P}}^b \triangleq \mathbb{E} \big\{\text{tr}(\Delta\mathbf{h}_{d,k}\Delta\mathbf{h}_{d,k}^H)\big\} = \frac{M\sigma^2}{p(\gamma+1)},
\end{equation}
 \vspace{-0.8cm}\begin{equation}\label{E-3}
\epsilon_{\bm{\mu}_k,\text{2P}}^b \triangleq \mathbb{E}\big\{\text{tr}(\Delta {\mathbf{x}}_k \Delta {\mathbf{x}}_k^H)  \big\} = \frac{\sigma^2}{p} \mathbb{E}\big\{\text{tr}\big((\hat{\mathbf{S}}^H_k\hat{\mathbf{S}}_k)^{-1}\big)   \big\} + \mathbb{E}\big\{\text{tr}({\hat{\mathbf{S}}_k}^{\dagger} \Delta{\mathbf{S}}_k{\bm{\mu}}_k {\bm{\mu}}_k^H \Delta{\mathbf{S}}^H_k ({\hat{\mathbf{S}}}_k^{\dagger})^H)  \big\}.
\end{equation}
\end{subequations}

\vspace{-0.3cm}
\noindent
 {Note that different from the case of $M\ge N$
 where the  MSE for estimating $\mathbf{h}_{d,k}$, i.e., $\epsilon_{d,k,\text{2P}}^a$, is related to $K$,
 $\epsilon_{d,k,\text{2P}}^b$ is  related to $\gamma$ here  since $\mathbf{h}_{d,k}$ is estimated  based on the received signals in $\gamma+1$ time slots.} 


 Let  $\mathbf{P}_{\bm{\mu}_k,1}^b  \triangleq  \frac{\sigma^2}{p} \mathbb{E}\big\{ \text{tr}\big((\hat{\mathbf{S}}^H_k\hat{\mathbf{S}}_k)^{-1}\big)   \big\}$ and  $\mathbf{P}_{\bm{\mu}_k,2}^b \triangleq \mathbb{E}\big\{\text{tr}({\hat{\mathbf{S}}_k}^{\dagger} \Delta{\mathbf{S}}_k{\bm{\mu}}_k {\bm{\mu}}_k^H \Delta{\mathbf{S}}^H_k ({\hat{\mathbf{S}}}_k^{\dagger})^H)  \big\}$ represent the two terms on the RHS  of \eqref{E-3}. In the sequel, we focus on further simplifying $\mathbf{P}_{\bm{\mu}_k,1}^b$ and $\mathbf{P}_{\bm{\mu}_k,2}^b$  when $M$ becomes large as the following proposition.
 \vspace{-0.3cm}
 \begin{proposition}\label{Pro3}\em
 {
$\mathbf{P}_{\bm{\mu}_k,1}^{\text{b}}$ and $\mathbf{P}_{\bm{\mu}_k,2}^{\text{b}}$ can be approximated by
$
\mathbf{P}_{\bm{\mu}_k,1}^b
\approx \frac{\sigma^2N(N+1)}{2pM(N+1)l^{\text{UI}}_1l^{\text{IB}}   + 2M\sigma^2}$ and
$
 \mathbf{P}_{\bm{\mu}_k,2}^b  \approx \frac{(\delta \gamma^2 +2\gamma + 2\delta-M)\sigma^2\sum_{n=1}^{(\gamma-1)M}\mathbb{E}\{\mu_{k,n}\mu_{k,n}^*\}}{p\delta (\gamma +1)^2(N+1)l^{\text{UI}}_1l^{\text{IB}} + (\delta \gamma^2 +2\gamma + 2\delta-M)\sigma^2}
+ \frac{(M\gamma^2+(3M-1)\gamma +\delta)\sigma^2 \sum_{n=(\gamma-1)M+1}^N\mathbb{E}\{\mu_{k,n}\mu_{k,n}^*\}}{p\delta (\gamma +1)^2(N+1)l^{\text{UI}}_1l^{\text{IB}} +(M\gamma^2+(3M-1)\gamma +\delta)\sigma^2 }
$ when $M$ becomes asymptotically large.}
\end{proposition}
\vspace{-0.5cm}
\begin{proof}
Please refer to Appendix C. 
\end{proof}

\vspace{-0.2cm}
By substituting the asymptotic values of $\mathbf{P}_{\bm{\mu}_k,1}^b $ and $\mathbf{P}_{\bm{\mu}_k,2}^b $  into \eqref{E-3},  the asymptotic MSE for estimating $\bm{\mu}_k$  can be expressed
as follows:
 \vspace{-0.3cm}\begin{equation}\label{epsilon}
\begin{aligned}
\epsilon_{\bm{\mu}_k,\text{2P}}^b
\approx & \frac{\sigma^2 N(N+1)}{2pM(N+1)l^{\text{UI}}_1l^{\text{IB}}   + 2M\sigma^2}
+ \frac{(\delta \gamma^2 +2\gamma + 2\delta-M)\sigma^2\sum_{n=1}^{(\gamma-1)M}\mathbb{E}\{\mu_{k,n}\mu_{k,n}^*\}}{p\delta (\gamma +1)^2(N+1)l^{\text{UI}}_1l^{\text{IB}} + (\delta \gamma^2 +2\gamma + 2\delta-M)\sigma^2}\\&
+ \frac{(M\gamma^2+(3M-1)\gamma +\delta)\sigma^2 \sum_{n=(\gamma-1)M+1}^N\mathbb{E}\{\mu_{k,n}\mu_{k,n}^*\}}{p\delta (\gamma +1)^2(N+1)l^{\text{UI}}_1l^{\text{IB}} +(M\gamma^2+(3M-1)\gamma +\delta)\sigma^2 }.
\end{aligned}
 \end{equation}

 \vspace{-0.3cm}\noindent
{Similar to \eqref{D-7}, we can see that as $p$, $l^{\text{UI}}_1$ and $l^{\text{IB}}$ decrease,  $\epsilon_{\bm{\mu}_k,\text{2P}}^b$
increases, and the first term on the RHS of \eqref{epsilon} dominates the other terms when $N$ is asymptotically large. Besides,
 different from the case of $M\ge N$, $\epsilon_{\bm{\mu}_k,\text{2P}}^b$ is related to $\gamma$ and $\delta$ in this case, which are determined by the quantitative relationship between $M$ and $N$.}

\vspace{-0.55cm}\subsection{Asymptotic MSE Comparison between 2PCE and 3PCE Strategies}\label{SEC4-3}
{
In this subsection, we compare the performance of the 2PCE and 3PCE strategies   in terms of the asymptotic MSE. For a fair comparison,  the asymptotic MSE performance of the 3PCE strategy is analyzed in a similar way as those in Section \ref{V-A} and Section \ref{V-B}, by assuming an LS estimator. The derived results are summarized in  Table \ref{Tab1},
\begin{table*}[t]
\vspace{-0.0em}
\setlength{\abovecaptionskip}{-0.1cm}
\setlength{\belowcaptionskip}{-0.6cm}
\caption{MSE performance of the 3PCE strategy in \cite{arXiv1912}}
\begin{center}
\begin{tabular}{|c|c|}
\hline
$\epsilon_{d,\text{3P}}$ & $\frac{M\sigma^2}{p}$ \\
\hline
$\epsilon_{r,1,\text{3P}}$ & $\frac{(1+K)M\sigma^2}{pK}$ \\
\hline
 $\epsilon_{r,k,\text{3P}}^{\text{a}}$ & $\frac{N(N-1)(1+K)\sigma^2}{KM\sigma^2+pKMNl^{\text{UI}}_1l^{\text{IB}}}+\frac{K(1+K)N\sigma^2}{K(K+N)M\sigma^2+pMNK^2l^{\text{UI}}_1l^{\text{IB}} }$\\
 &$+\frac{(K+N)M\sigma^2\mathbb{E}\{\mu_{k,1}\mu_{k,1}^*\}}{pN^2Kl^{\text{UI}}_1l^{\text{IB}}+(K+N)M\sigma^2}
+\frac{M\sigma^2\sum_{n=2}^N\mathbb{E}\{\mu_{k,n}\mu_{k,n}^*\}}{pN^2l^{\text{UI}}_1l^{\text{IB}}+M\sigma^2}$\\
\hline
$\epsilon_{\bm{\mu}_k,\text{3P}}^b$ &  $\frac{K(1+K)N\sigma^2}{pK^2NM l^{\text{UI}}_1l^{\text{IB}}+K(K+N)M\sigma^2}  + \frac{N(N-1)(1+K)\sigma^2}{pNMKl^{\text{UI}}_1l^{\text{IB}} +KM\sigma^2}+\frac{M\sigma^2\sum_{n=2}^{(\gamma-1)M} \mathbb{E}\{\mu_{k,n}\mu_{k,n}^*\}}{pNMl^{\text{UI}}_1l^{\text{IB}} +M\sigma^2} $\\
& $+
 \frac{(K+N)M\sigma^2 \mathbb{E}\{\mu_{k,1}\mu_{k,1}^*\}}{pKNMl^{\text{UI}}_1l^{\text{IB}}
 +(K+N)M\sigma^2} +
 \frac{M\sigma^2\sum_{n=(\gamma-1)M+1}^N\mathbb{E}\{\mu_{k,n}\mu_{k,n}^*\} }{pN \delta l^{\text{UI}}_1l^{\text{IB}}  + M\sigma^2}$\\
\hline
\end{tabular}
\label{Tab1}
\end{center}
\vspace{-1.0cm}
\end{table*}
%
 where $\epsilon_{d,\text{3P}}$ is the MSE for estimating  all the direct channels, $\epsilon_{r,1,\text{3P}}$ represents the MSE for estimating the reflected channel associated with $U_1$, $\epsilon_{\bm{\mu}_k,\text{3P}}^{\text{a}}$ and $\epsilon_{\bm{\mu}_k,\text{3P}}^b$ denote the asymptotic MSEs for estimating all the scaling factors in the cases of $M\ge N$ and $M < N$, respectively. Note that in the 3PCE strategy, $\{\mathbf{h}_{d,k}\}_{k=1}^K$, $\mathbf{H}_1$ and $\{\bm{\mu}_{k}\}_{k=2}^K$ are estimated successively, thus $\epsilon_{d,\text{3P}}$ is mainly due to  the channel noise,   $\epsilon_{r,1,\text{3P}}$ depends on the channel noise and the imperfect estimation of $\mathbf{h}_{d,1}$, while $\epsilon_{\bm{\mu}_k,\text{3P}}^{\text{a}}$ ($\epsilon_{\bm{\mu}_k,\text{3P}}^b$) is affected by the channel noise, the imperfect estimation of $\mathbf{h}_{d,1}$ and the imperfect estimation of $\mathbf{H}_1$.  In the sequel, we focus on the performance comparison between these two strategies and the main results are given in the following two theorems.
 \newtheorem{theorem}{Theorem}
 \vspace{-0.3cm}
 \begin{theorem}\label{PRO4}
 \em
 In the case of $M\ge N$, the asymptotic MSEs achieved by the 2PCE strategy for estimating all the direct and reflected channels are  lower than that achieved by the 3PCE strategy.
 \end{theorem}
 \vspace{-0.5cm}
\begin{proof}
Please refer to Appendix D. 
\end{proof}
\vspace{-0.5cm}
 \begin{theorem}\label{PRO5}
 \em
 In the case of $M<N$,  the MSE achieved by the 2PCE strategy for estimating all the direct channels is lower than that achieved by the 3PCE strategy, if $\gamma < K+2$. For estimating the reflected channels, the proposed 2PCE strategy  always outperforms the 3PCE strategy in terms of
 asymptotic MSE.
 \end{theorem}
 \vspace{-0.5cm}
\begin{proof}
Please refer to Appendix E. 
\end{proof}
}

\vspace{-0.7cm}\section{Simulation Results}
In this section, we present numerical results to verify the effectiveness  of the proposed 2PCE strategy. In our simulations, the distance-dependent path losses of ${\mathbf{h}}_{d,k}$'s, ${\mathbf{h}}_{r,k}$'s and $\mathbf{G}$ are modeled as
$l_{k,m}^{\text{UB}} =l_{0}\left({d_{k,m}}/{d_{0}}\right)^{-\alpha^{\text{UB}}}$, $l_{k,n}^{\text{UI}} =l_{0}\left({d_{k,n}}/{d_{0}}\right)^{-\alpha^{\text{UI}}}$
and $l^{\text{IB}}_{m,n} =l_{0}\left({d_{m,n}}/{d_{0}}\right)^{-\alpha^{\text{IB}}}$, respectively,
where $l_{0} = -30$ dB denotes the path loss at the reference distance $d_0 = 1$ meter (m), $d_{k,m}$, $d_{k,n}$ and $d_{m,n}$ represent the link distances from
 $U_k$ to the $m$-th BS antenna, from $U_k$ to the $n$-th reflecting element and from the $n$-th reflecting element to the $m$-th BS antenna, respectively, and $\alpha^{\text{UB}}$, $\alpha^{\text{UI}}$ and $\alpha^{\text{IB}}$ are path-loss exponents for the user-BS, user-IRS and IRS-BS links, respectively.
We assume that  the IRS is deployed to serve the users that suffer from severe signal attenuation
in the user-BS  link, thus the path-loss exponents are set to  $\alpha^{\text{UB}}= 5$, $\alpha^{\text{IB}}=2.2$ and $\alpha^{\text{UI}}=2.2$.  We consider a three-dimensional coordinate system  where the BS/IRS are deployed
 on the $x$-axis and $y$-$z$ plane, and we fix $N_y=4$. 
 The reference antenna/element at the BS/IRS are located at $(2\;\textrm{m}, 0, 0)$
and $(0, 45\;\textrm{m}, 2\;\textrm{m})$, and the antenna/element spacing are set to ${d_B} = {\mu}/{2}$ and ${d_I} = {\mu}/{8}$. The  power spectrum of the AWGN at the BS and the channel bandwidth are -169 dBm/Hz and 1 MHz.
 Besides,  we employ the normalized MSEs, i.e.,
$
\text{NMSE}_{d}=\frac{\sum_{k=1}^{K} \mathbb{E}\left\{\|\hat{\mathbf{h}}_{d,k}-\mathbf{h}_{d,k}\|^{2}\right\}}{\sum_{k=1}^{K} \mathbb{E}\left\{\|\mathbf{h}_{d,k}\|^{2}\right\}}$ and
$\text{NMSE}_r=\frac{\sum_{k=1}^{K} \mathbb{E}\left\{\|\hat{\mathbf{H}}_{k}-\mathbf{H}_{k}\|^{2}\right\}}{\sum_{k=1}^{K} \mathbb{E}\left\{\|\mathbf{H}_{k}\|^{2}\right\}}
$, as the performance metrics for estimating the direct and reflected channels,  respectively. Other system parameters are set as
follows unless otherwise specified: $K=4, N=32, M=40, \beta^{\text{IB}} = 3$ dB, $\beta^{\text{UB}} = \beta^{\text{UI}}=0$, $r_d=r_r=r_{r,k}=0$. All the results are averaged over 2000 independent channel realizations.
\subsubsection{Impact of the transmit power, $p$}

\begin{figure}[b]
\vspace{-1.3cm}
\setlength{\belowcaptionskip}{-1cm}
\renewcommand{\captionfont}{\small}
\begin{minipage}[t]{0.5\textwidth}
\centering
\includegraphics[scale=.48]{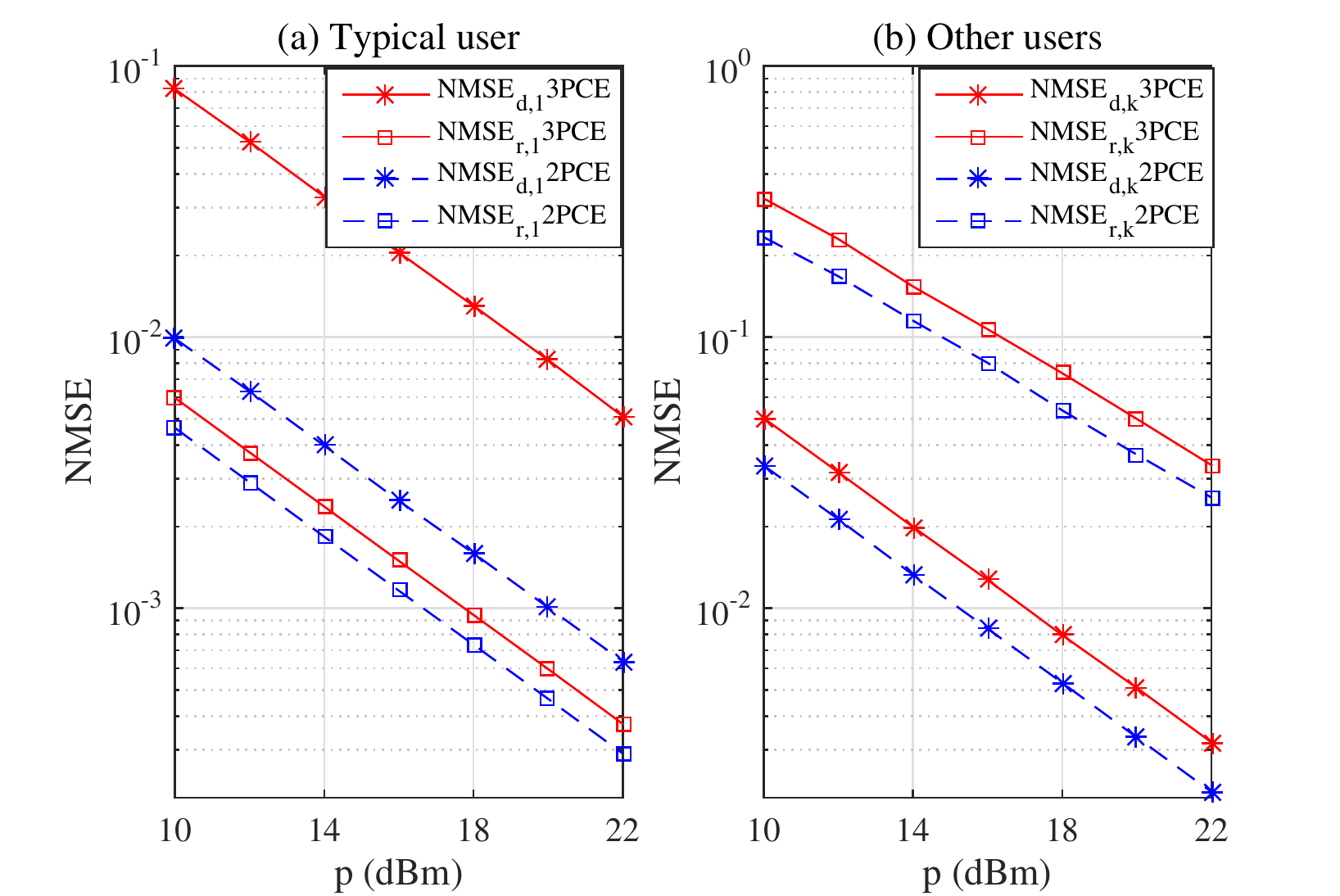}
\caption{NMSE performance versus the  transmit power, $p$. }
\label{LSnmse}
\normalsize
\end{minipage}
\;
\begin{minipage}[t]{0.5\textwidth}
\centering
\includegraphics[scale=.48]{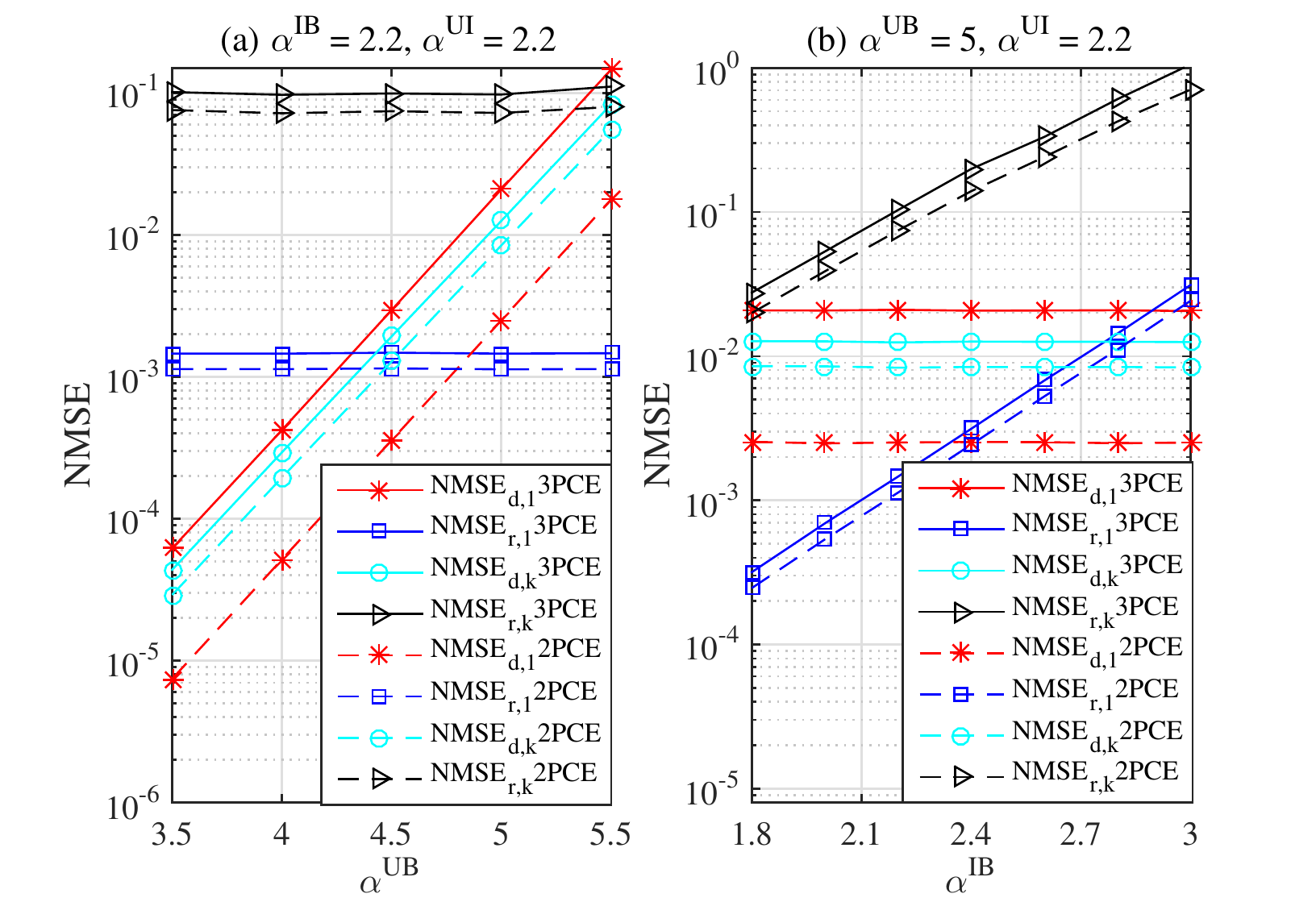}
 \caption{ NMSE performance  versus the path-loss  exponents, $\alpha^{\text{UB}}$ and $\alpha^{\text{IB}}$.}
 \label{LSnmsePathloss}
\normalsize
\end{minipage}
\end{figure}
First, we investigate in Fig. \ref{LSnmse}  the NMSE performance of the proposed 2PCE strategy versus the transmit power $p$, and the performance of the 3PCE strategy \cite{arXiv1912} is regarded as the performance benchmark.
 Let $\text{NMSE}_{d,1}$ ($\text{NMSE}_{r,1}$) and $\text{NMSE}_{d,k}$ ($\text{NMSE}_{r,k}$) represent the $\text{NMSE}$s for estimating $\mathbf{h}_{d,1}$ ($\mathbf{H}_1$) and $\{\mathbf{h}_{d,k} \}_{k=2}^K$ ($\{\mathbf{H}_{k} \}_{k=2}^K$), respectively.
First, it is observed that the NMSE performance of both strategies improves with the increasing of $p$, which is expected since larger $p$ implies higher channel training SNR.
 Besides, we can observe from Fig. \ref{LSnmse} (a) that for the typical user, the 2PCE strategy can achieve much better NMSE performance than the 3PCE strategy when estimating the direct channel $\mathbf{h}_{d,1}$, {which is mainly due to the fact that  $\mathbf{h}_{d,1}$ is obtained based on the received signals in a larger total number of time slots as compared to the 3PCE strategy. } 
 Moreover, the  2PCE strategy outperforms the 3PCE strategy when estimating $\mathbf{H}_1$, which is because the estimation error of $\mathbf{h}_{d,1}$ will not affect the estimation of $\mathbf{H}_1$ in the proposed 2PCE strategy.
Due to similar reasons, the NMSE performance achieved by the proposed 2PCE strategy for the estimated channels associated with the other users  is also better than that achieved by the 3PCE strategy, as can be observed from Fig. \ref{LSnmse} (b). Specifically, to achieve the same NMSE performance, the transmit power required by the proposed 2PCE strategy is about 2 dB less than that required by the 3PCE strategy. Finally, it is noteworthy that this performance gain is available for all training SNR regimes.

\subsubsection{Impact of the path-loss coefficients, $\alpha^{\text{UB}}$ and  $\alpha^{\text{IB}}$}
{Fig. \ref{LSnmsePathloss} shows the NMSE performance versus the path-loss exponents of the user-BS link $\alpha^{\text{UB}}$ and the  IRS-BS link  $\alpha^{\text{IB}}$, where the user transmit power is set to $p=16$ dBm.
Besides, we fix $\alpha^{\text{IB}}=2.2$ and $\alpha^{\text{UI}}=2.2$ in Fig. \ref{LSnmsePathloss} (a), while in Fig. \ref{LSnmsePathloss} (b), we fix $\alpha^{\text{UB}}=5$ and $\alpha^{\text{UI}}=2.2$. It is observed that the proposed 2PCE strategy outperforms the 3PCE strategy, similar to the results in Fig. \ref{LSnmse}. Besides,
from Fig.  \ref{LSnmsePathloss} (a), we observe that
the NMSEs achieved by the 2PCE and 3PCE strategies for estimating the direct channels decrease with the increasing of $\alpha^{\text{UB}}$, while those for estimating the reflected channels remain unchanged with different values of $\alpha^{\text{UB}}$. This is because when $\alpha^{\text{UB}}$ increases, the path loss of each user-BS link becomes more severe and  it becomes more difficult to accurately estimate the direct channels.  Similarly,
 from Fig. \ref{LSnmsePathloss} (b), it can be seen that the NMSEs achieved by these two strategies for estimating the reflected channels decrease as $\alpha^{\text{IB}}$ increases, while those for estimating the direct channels are not affected by  the value of $\alpha^{\text{IB}}$.
 For the 3PCE strategy, this is because the estimation of the direct channels is conducted when the IRS is switched off  and thus not affected by the estimation of the reflected channels. While for the proposed 2PCE strategy, the results in Fig. \ref{LSnmsePathloss} (b) are consistent with the analysis in Section IV (see \eqref{E2k}), i.e., although the estimation accuracy of the direct channels associated with the other users in Phase II depends on the knowledge of $\mathbf{H}_1$, the MSE performance is not affected by the estimation accuracy of $\mathbf{H}_1$.
Note that by fixing  $\alpha^{\text{UB}}=5$, $\alpha^{\text{IB}}=2.2$ and
 investigating the impact of $\alpha^{\text{UI}}$ on
the NMSE performance, similar results as those in Fig. \ref{LSnmsePathloss} (b) can be obtained and therefore the details are omitted for
brevity.}

\begin{figure}[b]
\vspace{-1.3cm}
\setlength{\belowcaptionskip}{-1cm}
\renewcommand{\captionfont}{\small}
\begin{minipage}[t]{0.5\textwidth}
\centering
\includegraphics[scale=.48]{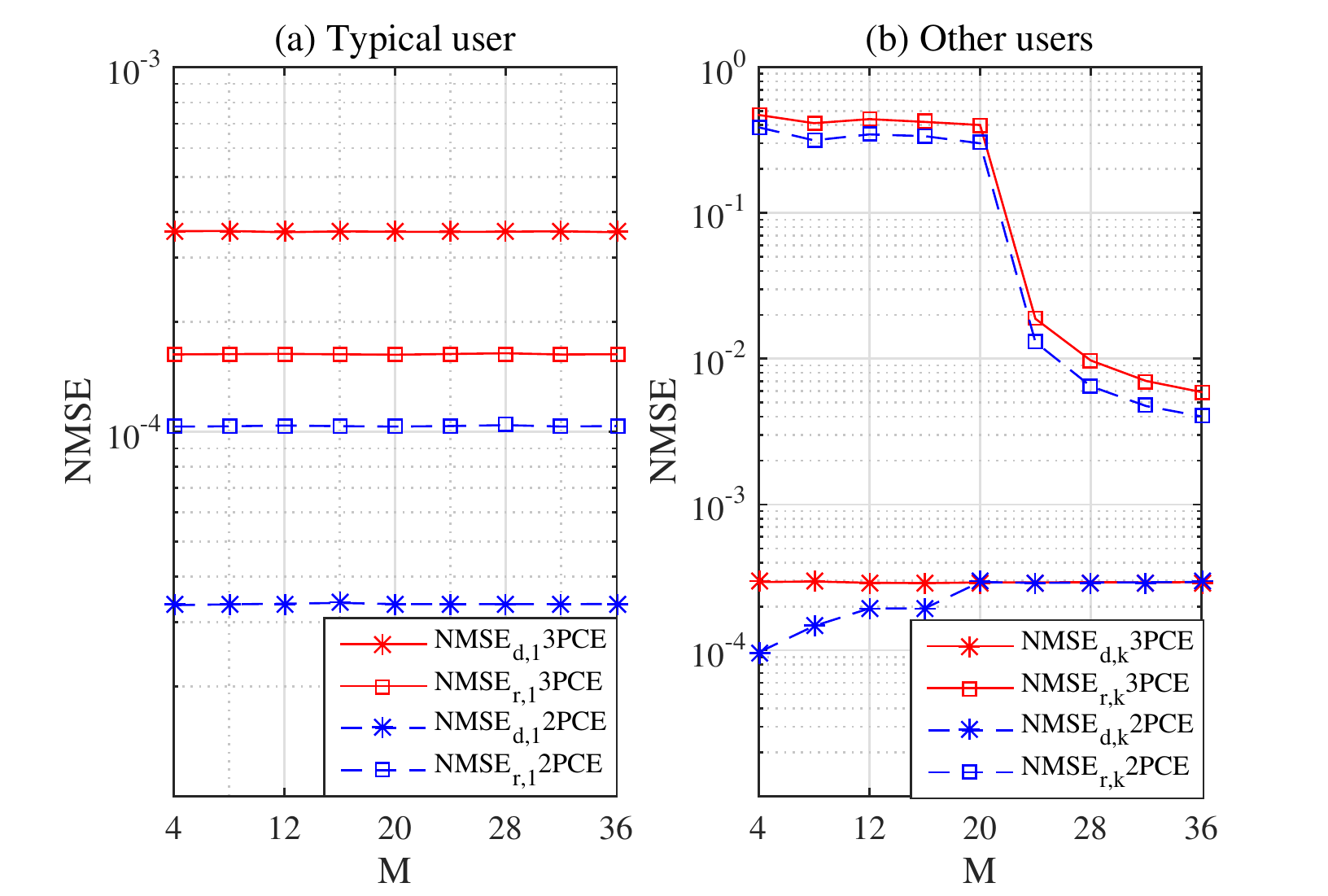}
\caption{NMSE performance  versus the number of BS antennas, $M$. }
\label{LSnmseM}
\normalsize
\end{minipage}
\;
\begin{minipage}[t]{0.5\textwidth}
\centering
\includegraphics[scale=.48]{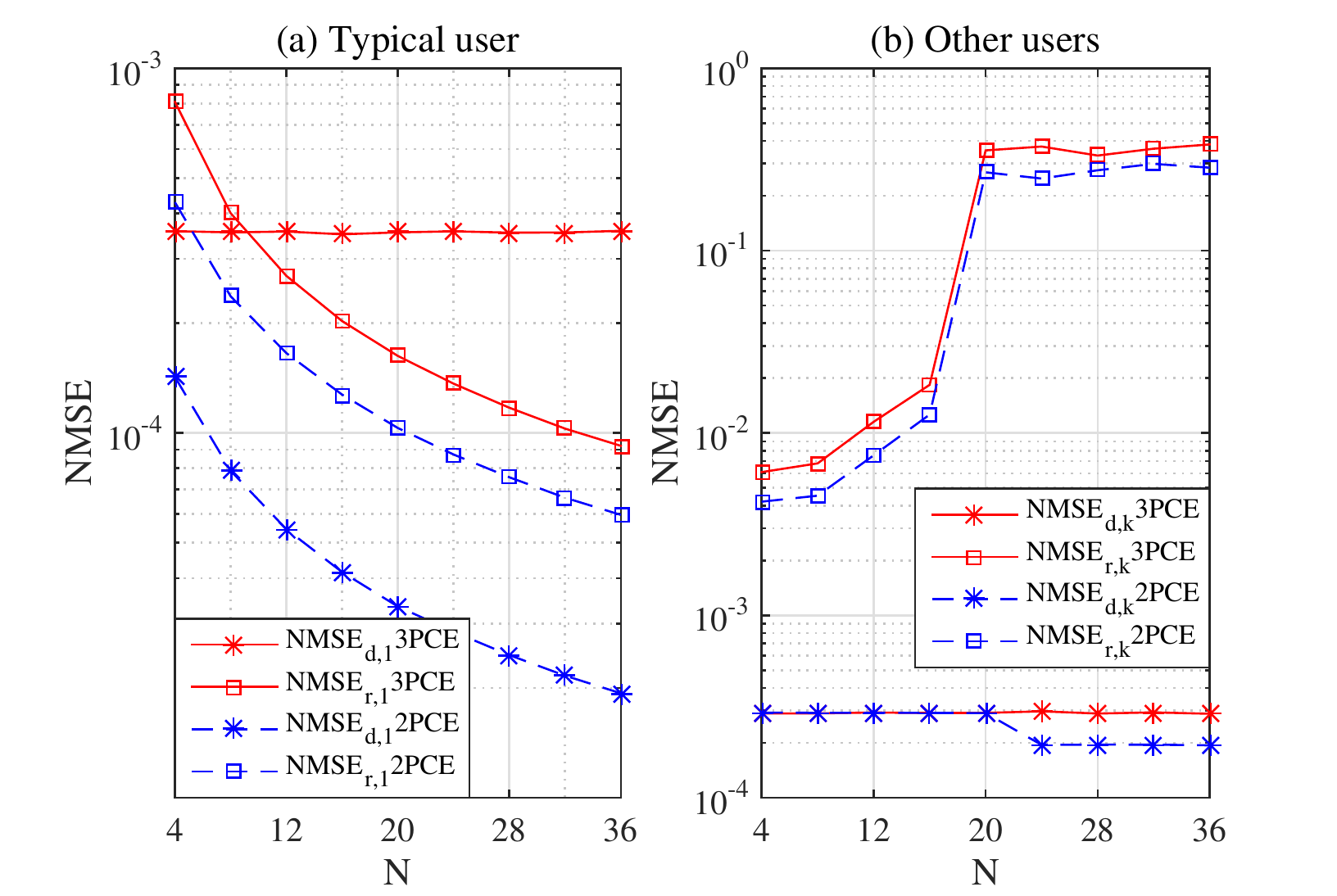}
 \caption{ NMSE performance  versus the number of reflecting elements, $N$.}
 \label{LSnmseN}
\normalsize
\end{minipage}
\end{figure}

\subsubsection{Impact of the number of BS antennas $M$}
 {In Fig. \ref{LSnmseM}, we investigate the impact of the number of BS antennas $M$ on the NMSE performance, where we fix $N=20$, $p=24$ dBm and  $K=2$.
 First, we observe that the proposed 2PCE strategy outperforms the 3PCE strategy, as in Figs. \ref{LSnmse} and \ref{LSnmsePathloss}.
 Second, from Fig. \ref{LSnmseM} (a), we observe that  the NMSE performance  is almost invariant with different values of $M$. This is because  the MSEs for estimating $\mathbf{h}_{d,1}$ and $\mathbf{H}_1$ in the 2PCE strategy, i.e., $\epsilon_{d,1,\text{2P}}$ and $\epsilon_{r,1,\text{2P}}$, increase with $M$
  (according to \eqref{E21}), thus the corresponding NMSEs, i.e., $\text{NMSE}_{d,1}$ and $\text{NMSE}_{r,1}$, are not related to $M$ due to the normalization. Then, one can see from  Fig. \ref{LSnmseM} (b) that in the case of $ M<N$, the $\text{NMSE}_{d,k}$ performance of the 2PCE strategy deteriorates with the increasing $M$ because the asymptotic MSE for estimating $\mathbf{h}_{d,k},k\neq 1$, i.e., $\epsilon_{d,k,\text{2P}}^b$, is inversely proportional to $\gamma +1 $ according to \eqref{L6-13} and the value of $\gamma$
  decreases when $M$ becomes larger. In the case of $M\ge N$, since $\epsilon_{d,k,\text{2P}}^a$ is linearly proportional to $M$ according to \eqref{E2k}, the $\text{NMSE}_{d,k}$ achieved by the 2PCE strategy is not related to $M$ because of the normalization, and it is similar  to that achieved by the 3PCE strategy. Besides, for the estimation of the reflected channels associated with the other users, the NMSE performance achieved by the 2PCE strategy does not change much when $M <N$, while it
 improves with the increasing of $M$ when $M\ge N$. The main reason is that  in the proposed   2PCE strategy, the
inverses of the $2M\times (M+N)$ matrix ${\hat{\mathbf{V}}}^{\text{II}}$ and the $(\gamma+1)M \times (M+N)$ matrix ${\hat{\mathbf{Q}}}^{\text{II}}_k$ are required to estimate $\mathbf{H}_k, k \neq 1$, for the cases of  $M\ge N$ and $M<N$, respectively. Let $\varrho_V = \frac{2M}{M+N}$ and $\varrho_Q = \frac{(\gamma+1)M}{M+N}$ denote the ratios of the number of rows to the number of columns  of the matrices ${\hat{\mathbf{V}}}^{\text{II}}$ and ${\hat{\mathbf{Q}}}^{\text{II}}_k$, respectively. Then, we can see that $\varrho_V$ increases with $M$ when $M\ge N$ and   $\hat{\mathbf{V}}^{\text{II}}$ will become less ill-conditioned as $M$ increases,  which causes less performance loss when estimating $\mathbf{H}_k$. However, when $M<N$,  $\varrho_M$ satisfies $\varrho_Q \ge 1$ and $\varrho_Q \approx 1$  with different values of $M$, therefore $\text{NMSE}_{r,k}$ is almost invariant with $M$ in this case.}
%
%
\subsubsection{Impact of the number of reflecting elements $N$}
{In Fig. \ref{LSnmseN}, we investigate the NMSE performance versus the number of reflecting elements $N$ with fixed $M=20$, $p=24$ dBm and  $K=2$. Similar to the results presented above, we can observe that the NMSE performance of the 2PCE strategy is superior to that of the 3PCE strategy.
From Fig. \ref{LSnmseN} (a), it can be seen that the NMSE performance achieved by the 2PCE strategy for estimating $\mathbf{h}_{d,1}$ and ${\mathbf{H}}_1$ improves with $N$, which is mainly because both $\mathbf{h}_{d,1}$ and $\mathbf{H}_1$ are estimated based on the received signals in $N+1$ time slots. Fig. \ref{LSnmseN} (b) shows that $\text{NMSE}_{d,k}$ remains unchanged with varying $N$ when $M\ge N$. This is due to the fact that  $\epsilon_{d,k,\text{2P}}^a$ is not related to $N$  according to \eqref{E2k}. However, when $M < N$, $\text{NMSE}_{d,k}$ decreases with the increasing of  $N$,
which is consistent with analysis in \eqref{L6-13}.
 Furthermore,  focusing on  the estimation of $\mathbf{H}_{k},k\neq 1$, we can see that  the NMSE performance achieved by the 2PCE strategy deteriorates with the increasing of $N$ due to the reduced value of $\varrho_V$ when $M\ge N$ (similar to that in Fig. \ref{LSnmseM} (b)), while the performance  is nearly invariant when $M<N$ since $\varrho_Q \approx 1$  holds under different values of $N$.

 }

\begin{figure}[b]
\vspace{-1.2cm}
\setlength{\belowcaptionskip}{-1cm}
\renewcommand{\captionfont}{\small}
\begin{minipage}[t]{0.5\textwidth}
\centering
\includegraphics[scale=.48]{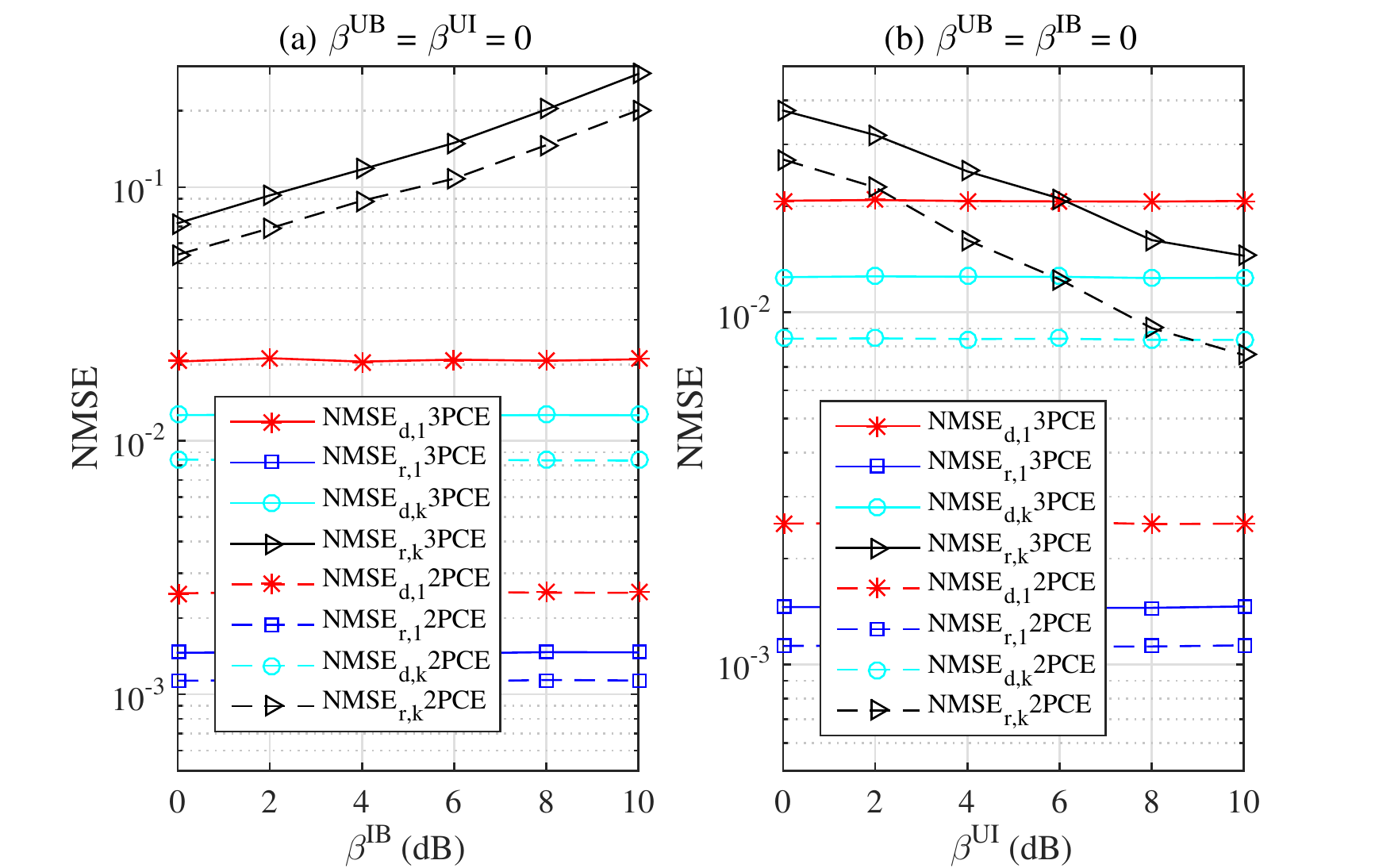}
\caption{NMSE performance versus the Rician factors, $\beta^{\text{IB}}$ and $\beta^{\text{UI}}$. }
\label{nmseK-Factor}
\normalsize
\end{minipage}
\;
\begin{minipage}[t]{0.5\textwidth}
\centering
\includegraphics[scale=.48]{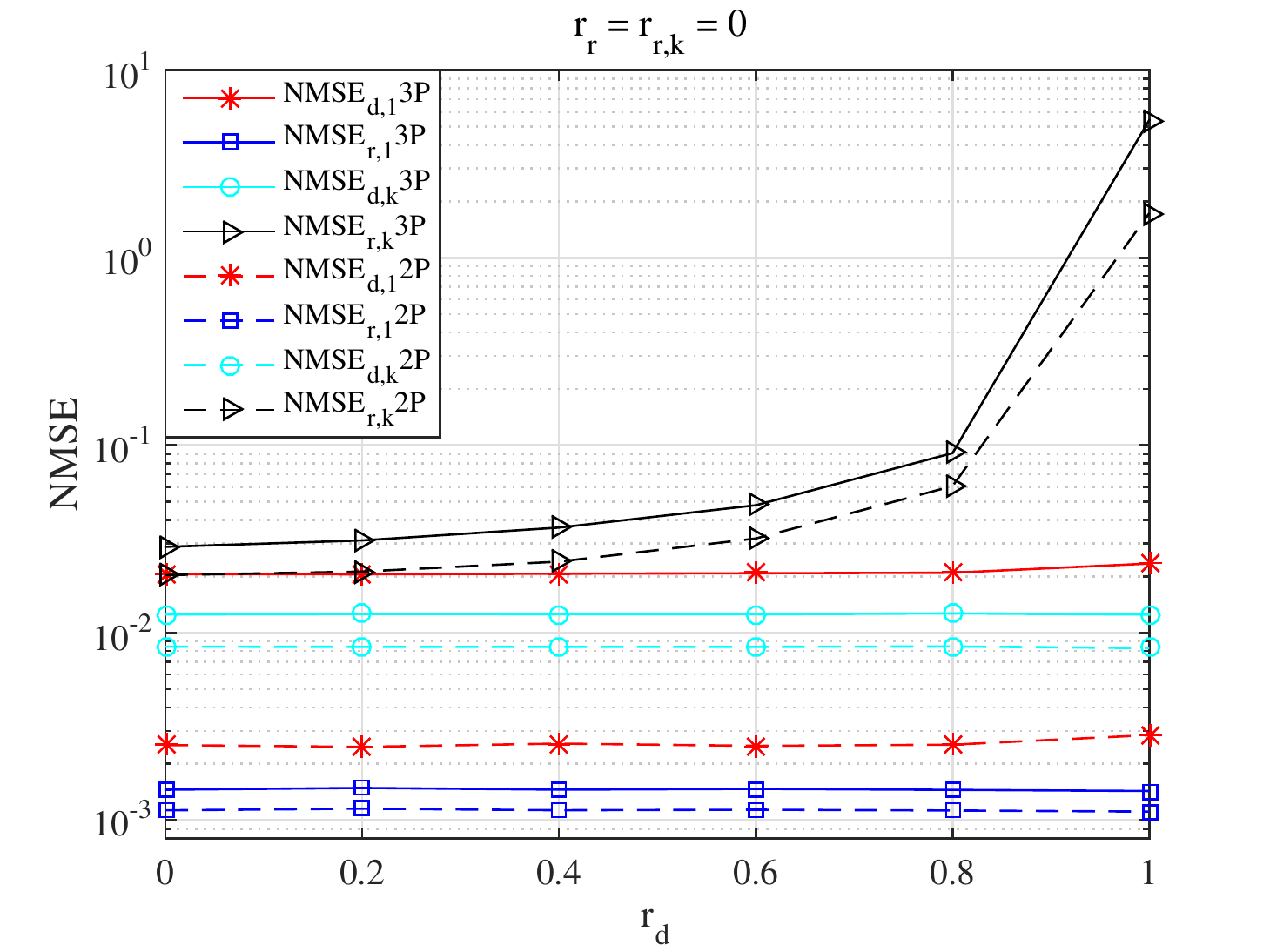}  
\caption{NMSE performance versus the correlation
coefficient, $r_d$. }
\label{nmseR}
\normalsize
\end{minipage}
\end{figure}

\subsubsection{Impact of the Rician factors, $\beta^{\text{IB}}$ and $\beta^{\text{UI}}$}
{
 Fig. \ref{nmseK-Factor} investigates the NMSE performance versus the Rician factors of the IRS-BS link $\beta^{\text{IB}}$ and the user-IRS link $\beta^{\text{UI}}$ with fixed $p = 16$ dBm. First, in  Fig. \ref{nmseK-Factor} (a), we fix  $\beta^{\text{UI}}=\beta^{\text{UB}}=0$ and it can be seen that the NMSEs achieved by the 2PCE and 3PCE strategies for estimating the reflected channels $\mathbf{H}_k,k\neq 1$, increase with the increasing of $\beta^{\text{IB}}$, while the $\text{NMSE}_{d,1}$, $\text{NMSE}_{d,k}$ and  $\text{NMSE}_{r,1}$ performance  do not change much with varying $\beta^{\text{IB}}$. The main reason is that as $\beta^{\text{IB}}$ increases, the IRS-BS channel becomes more
 deterministic and the rank of the channel  decreases in general (according to the channel model in \eqref{eq7}), thus the channel matrix $\mathbf{H}_1$
  becomes more ill-conditioned, which
  will affect the estimation of $\mathbf{H}_k$
   when calculating the pseudo-inverse of $\hat{\mathbf{V}}^{\text{II}}$ in the proposed 2PCE strategy (or $\hat{\mathbf{H}}_1$ in the 3PCE strategy).
Second, in Fig. \ref{nmseK-Factor} (b),
we fix  $\beta^{\text{IB}}=0$ and $\beta^{\text{UB}}=0$ and it can be observed that
the NMSE performance achieved by these two strategies improves with the increasing of $\beta^{\text{UI}}$, which is expected since the channel matrix $\mathbf{H}_1$ becomes less ill-conditioned as $\beta^{\text{UI}}$ increases.\footnote{ According to the user-IRS channel model in \eqref{eq7},  as $\beta^{\text{UI}}$ increases, the user-IRS channel associated with $U_k$, i.e., $\mathbf{h}_{r,k}$, becomes more deterministic and more like its LoS element ${\bar{\mathbf{z}}}_{r,k}$ whose entries have the same amplitude according to \eqref{eq7} and \eqref{eq4}.
As a result,  the reflected channel $\mathbf{H}_k$, whose $n$-th column is the scalar-vector multiplication of the $n$-th entry of $\mathbf{h}_{r,k}$ and the $n$-th column of the IRS-BS channel $\mathbf{G}$ (see the reflected channel model given in \eqref{A-11}), is less ill-conditioned if $\beta^{\text{UI}}$ is larger. } Furthermore, one can observe that the benefit brought by increasing $\beta^{\text{UI}}$ is more pronounced on the proposed 2PCE strategy than that on the 3PCE strategy.}

\subsubsection{Impact of the correlation coefficient, $r_d$}
  Finally,  we plot in Fig. \ref{nmseR} the NMSE performance achieved by the 2PCE and 3PCE strategies versus the correlation coefficients $r_d$, where we assume $r_r = r_{r,k} = 0$ and $p = 16$ dBm. As can be observed,  only the NMSE performance for estimating $\{\mathbf{H}_k\}_{k=2}^K$ is affected by  the increasing of $r_d$, while the impact of $r_d$ on $\text{NMSE}_{d,1}$, $\text{NMSE}_{d,k}$ and  $\text{NMSE}_{r,1}$ is not that significant. This is due to the fact that as  $r_d$ increases, the IRS-BS channel $\mathbf{G}$ becomes more correlated and thus the reflected channel $\mathbf{H}_1$ becomes more ill-conditioned, which could result in  NMSE performance loss. 
  Besides, it is noteworthy that the proposed 2PCE strategy outperforms the 3PCE strategy for all values of $r_d$.
  {Similar result as in Fig. \ref{nmseR} can be observed when fixing $r_d = r_{r,k} = 0$ and investigating the impact of  $r_{r}$ on
the NMSE performance, while the NMSE performance is shown to be invariant with $r_{r,k}$ by assuming $r_{d} = r_r = 0$ and studying the impact of $r_{r,k}$,     hence their details are not shown here.

  }

\vspace{-0.3cm}\section{Conclusion}
This work studied the uplink channel estimation problem for an IRS-aided  multiuser SIMO system and proposed a novel  2PCE strategy.
By reducing the channel estimation phases and estimating the direct and reflected channels associated with each user simultaneously, the proposed 2PCE strategy is able to alleviate the negative
effects caused by error propagation and enhance the channel estimation performance without increasing the channel training overhead. In addition, the asymptotic MSE of the proposed 2PCE strategy is analyzed when the
LS channel estimation method is employed.
Both theoretical analysis and simulation results  demonstrated that the proposed 2PCE strategy can achieve better performance than the state-of-the-art 3PCE strategy \cite{arXiv1912} under various system parameters.


\end{spacing}
\begin{spacing}{1.45}
\ifCLASSOPTIONcaptionsoff
  \newpage
\fi

\newcounter{app}
\newcommand{\append}[1]{ \stepcounter{app}\centerline{\rm {Appendix \Alph{app}}{#1}}}

%
\vspace{0.2cm}
{\append{ \;\;Necessary propositions and lemmas}\label{AppenA}}
\vspace{-0.3cm}
\begin{proposition}\label{PRO1}
\rm
For any random  matrix $\mathbf{X}\in \mathbf{C}^{p\times n}$ with $p\le n$, if $\mathbb{E}\{\mathbf{X}{\mathbf{X}}^H\}$ is a diagonal matrix, i.e., the columns of $\mathbf{X}$ are independent with each other, then $\mathbb{E}\big\{(\mathbf{X} \mathbf{X}^H)^{-1} \big\} = \big( \mathbb{E}\{\mathbf{X}{\mathbf{X}}^H\}\big)^{-1}$ holds as $n \rightarrow \infty$.
\end{proposition}

\vspace{-0.5cm}
\begin{proof}
\rm
 First, it is easy to see that if $\mathbb{E}\{\mathbf{X}{\mathbf{X}}^H\}$ is a diagonal matrix, then each column of $\mathbf{X}$ is independent with each other.
 By defining  $\mathbf{C}_{\mathbf{X}} = \mathbb{E}\{\mathbf{X}{\mathbf{X}}^H\}$ and $\mathbf{Y} = \mathbf{C}_{\mathbf{X}}^{-\frac{1}{2}}\mathbf{X}$, we can see that  the random matrix $\mathbf{Y} = [\mathbf{y}_1,\cdots,\mathbf{y}_n]$  consists of $n$ independent random vectors $\{ \mathbf{y}_i\}_{i=1}^n$, which are  from the same distribution with covariance matrix $\mathbb{E}\{\mathbf{y}\mathbf{y}^H\}=\frac{1}{n}\mathbf{I}$.
Furthermore, defining  $\hat{\mathbf{C}}_{\mathbf{y},n}$ as the sample covariance matrix of  $\mathbf{y}$ defined over $n$ samples, i.e.,
$
\hat{\mathbf{C}}_{\mathbf{y},n} = \frac{1}{n}\sum\limits_{i=1}^n \mathbf{y}_i \mathbf{y}_i^H = \frac{1}{n}\mathbf{Y}\mathbf{Y}^H
$, then $\hat{\mathbf{C}}_{\mathbf{y},n}$ approximates the true  covariance matrix $\mathbb{E}\{\mathbf{y}\mathbf{y}^H\}$ as $n\rightarrow \infty$, and equivalently  we have $\mathbf{Y}\mathbf{Y}^H \rightarrow \mathbf{I}$ as $n \rightarrow \infty$ \cite{Livan2018,GIORGIO,COVARIANCE2009}.  Therefore, according to the relationship between $\mathbf{X}$ and $\mathbf{Y}$, i.e.,  $\mathbf{Y} = \mathbf{C}_{\mathbf{X}}^{-\frac{1}{2}}\mathbf{X}$, it follows that
$
 \mathbf{X}\mathbf{X}^H  = \mathbf{C}_{\mathbf{X}}.
$
Since $\mathbf{C}_{\mathbf{X}}$ is a constant diagonal matrix, we have $(\mathbf{X}\mathbf{X}^H )^{-1}\rightarrow \mathbf{C}_{\mathbf{X}}^{-1}$ as $n \rightarrow \infty$  and hence $\mathbb{E}\big\{(\mathbf{X}\mathbf{X}^H )^{-1}\big\} = \mathbf{C}_{\mathbf{X}}^{-1}$ holds when $n \rightarrow \infty$. This completes the proof.

\end{proof}
\vspace{-0.8cm}
\begin{lemma}\label{lemma1}
\rm
\cite{Bernstein2006}
Given a $2\times 2$ block matrix $\mathbf{M} = [\mathbf{A}, \mathbf{B};\mathbf{C}, \mathbf{D}]$,
if both $\mathbf{A}$ and $\mathbf{D}$ are square and both $\mathbf{A}$ and $\mathbf{D}-\mathbf{C A}^{-1} \mathbf{B}$ are invertible, then   the inverse of $\mathbf{M}$ can be expressed as follows:
 \vspace{-0.3cm}\begin{equation}
\left[\begin{array}{cc}
\mathbf{A} & \mathbf{B} \\
\mathbf{C} & \mathbf{D}
\end{array}\right]^{-1}=\left[\begin{array}{cc}
\mathbf{A}^{-1}+\mathbf{A}^{-1} \mathbf{B}\left(\mathbf{D}-\mathbf{C A}^{-1} \mathbf{B}\right)^{-1} \mathbf{C A}^{-1} & -\mathbf{A}^{-1} \mathbf{B}\left(\mathbf{D}-\mathbf{C A}^{-1} \mathbf{B}\right)^{-1} \\
-\left(\mathbf{D}-\mathbf{C A}^{-1} \mathbf{B}\right)^{-1} \mathbf{C A}^{-1} & \left(\mathbf{D}-\mathbf{C A}^{-1} \mathbf{B}\right)^{-1}
\end{array}\right].
\end{equation}
\end{lemma}
\vspace{-0.8cm}
\begin{lemma}\label{lemma2}
\rm
\cite{Hartwing1976}
Given a $2\times 2$ block matrix $\mathbf{M} = [\mathbf{A}, \mathbf{B};\mathbf{C}, \mathbf{D}]$,
if $
(\mathbf{I} - \mathbf{A}\mathbf{A}^{\dagger})\mathbf{B} = \mathbf{0}, \mathbf{C}(\mathbf{I} - \mathbf{A}^{\dagger}\mathbf{A}) = \mathbf{0}
$ and $\mathbf{D}-\mathbf{C} \mathbf{A}^{\dagger} \mathbf{B} = \mathbf{0}
$, then  the pseudo-inverse of $\mathbf{M}$ is given by
 \vspace{-0.3cm}\begin{equation}
\mathbf{M}^{\dagger}=\left[\begin{array}{cc}
\left(\mathbf{I}-\mathbf{K}_B \tilde{\mathbf{K}}_B^{-1} \mathbf{K}_B^{H}\right) \mathbf{A}^{\dagger}\left(\mathbf{I}-\mathbf{K}_C^{H} \tilde{\mathbf{K}}_C^{-1} \mathbf{K}_C\right) & \left(\mathbf{I}-\mathbf{ K}_B \tilde{\mathbf{K}}_B^{-1} \mathbf{K}_B^{H}\right) \mathbf{A}^{\dagger} \mathbf{K}_C^{H} \tilde{\mathbf{K}}_C^{-1} \\
\tilde{\mathbf{K}}_B^{-1} \mathbf{K}_B^{H} \mathbf{A}^{\dagger}\left(\mathbf{I}-\mathbf{K}_C^{H} \tilde{\mathbf{K}}_C^{-1} \mathbf{K}_C\right) & \tilde{\mathbf{K}}_B^{-1} \mathbf{K}_B^{H} \mathbf{A}^{\dagger} \mathbf{K}_C^{H} \tilde{\mathbf{K}}_C^{-1}
\end{array}\right],
 \end{equation}

 \vspace{-0.2cm}\noindent
where
$
\mathbf{K}_B=\mathbf{A}^{\dagger} \mathbf{B}$, $\mathbf{K}_C=\mathbf{C} \mathbf{A}^{\dagger}$, $\tilde{\mathbf{K}}_B = \mathbf{I}+\mathbf{K}_B^H \mathbf{K}_B$ and
$\tilde{\mathbf{K}}_C = \mathbf{I}+\mathbf{K}_C\mathbf{K}_C^H$.
\end{lemma}
\vspace{-0.6cm}
\begin{lemma}\label{lemma3}
\rm
For any square matrix $\mathbf{X}\in {\mathbb{C}}^{N\times N}$, we have $\mathbb{E}\big\{\text{tr}(\mathbf{X})\big\} = \text{tr}\big( \mathbb{E}\{\mathbf{X}\}\big)$.
\end{lemma}
\vspace{-0.4cm}
\begin{proof}\rm
Since $
\mathbb{E}\big\{\text{tr}(\mathbf{X})\big\}
= \mathbb{E}\{\sum_{l=1}^L\sum_{n=1}^N\mathbf{X}_{n,n}\},
$ and $\text{tr}\big( \mathbb{E}\{\mathbf{X}\}\big) = \sum_{n=1}^N\mathbb{E}\{\mathbf{X}_{n,n}\}$, we can easily prove $\mathbb{E}\big\{\text{tr}(\mathbf{X})\big\} = \text{tr}\big( \mathbb{E}\{\mathbf{X}\}\big)$ as the expected value of the sum of some random variables is equal to the sum of their individual expected values regardless of whether they are independent.
\end{proof}
\vspace{-0.5cm}
\begin{lemma}\label{lemma4}
\rm
\cite{Meyer2000} Any product of conformable
matrices can be permuted cyclically without altering the trace of the product, i.e.,
$
\text{tr}(\mathbf{ABC}) = \text{tr}(\mathbf{BCA}) = \text{tr}(\mathbf{CBA}).
$
\end{lemma}

\vspace{-0.2cm}
{\append{\;\; Proof of Proposition \ref{Pro2}}\label{AppenB}}

First, by resorting to Lemma \ref{lemma3}, $\mathbf{P}_{\bm{\mu}_k,1}^{\text{a}}$ can be transformed into $\mathbf{P}_{\bm{\mu}_k,1}^{\text{a}} {=}  \frac{\sigma^2}{2p(K-1)}\text{tr}\big( \mathbb{E} \big\{(\hat{{\mathbf{H}}}_1^H{\hat{\mathbf{H}}}_1)^{-1}\big\}\big)$.
Then, we have $
\mathbb{E}\{\hat{{\mathbf{H}}}_1^H{\hat{\mathbf{H}}}\} \overset{\text{(a)}}{=} \mathbb{E}\{\mathbf{H}_1^H\mathbf{H}_1\}+
\mathbb{E}\{\Delta\mathbf{H}_1^H\Delta\mathbf{H}_1\} \overset{\text{(b)}}{=} \frac{M(p(N+1) l^{\text{UI}}_1l^{\text{IB}} + \sigma^2)}{p(N+1)}\mathbf{I}_N
$,
 where $(\text{a})$ is due to the fact that the elements in $\mathbf{H}_1$ and $\Delta\mathbf{H}_1$  are independent with each other, and $(\text{b})$ is obtained based on  \eqref{A-8} and the  reflected channel model given in \eqref{A-11}, thus $\mathbb{E}\{\hat{{\mathbf{H}}}_1^H{\hat{\mathbf{H}}}\}$ is a diagonal matrix.
 Besides, when $M$ becomes asymptotically large, we have $\mathbb{E}\big\{(\hat{{\mathbf{H}}}_1^H{\hat{\mathbf{H}}}_1)^{-1}\big\}
\approx \big( \mathbb{E}\{\hat{{\mathbf{H}}}_1^H{\hat{\mathbf{H}}}\} \big)^{-1}
= \frac{p(N+1)}{M(p(N+1) l^{\text{UI}}_1l^{\text{IB}} + \sigma^2)}\mathbf{I}_N$ according to Proposition \ref{PRO1}. Therefore, $ \mathbf{P}_{\bm{\mu}_k,1}^{\text{a}}$ can be approximated by
 $\mathbf{P}_{\bm{\mu}_k,1}^{\text{a}} \approx \frac{\sigma^2(N+1)N}{2(K-1)M(p(N+1) l^{\text{UI}}_1l^{\text{IB}} + \sigma^2)}$.
  Second, $\mathbf{P}_{\bm{\mu}_k,2}^{\text{a}}$ can be equivalently rewritten as
 \vspace{-0.3cm}\begin{equation}\label{Pmuk2a}
\mathbf{P}_{\bm{\mu}_k,2}^{\text{a}}
\overset{\text{(a)}}{=}  \mathbb{E}\Big\{\text{tr}\big( \bm{\mu}_k\bm{\mu}_k^H\Delta\mathbf{H}_1^H(\hat{\mathbf{H}}_1^H)^{\dagger}\hat{\mathbf{H}}_1^{\dagger}\Delta\mathbf{H}_1 \big)\Big\}
\overset{\text{(b)}}{=} \text{tr}\Big( \mathbb{E}\big\{ \bm{\mu}_k\bm{\mu}_k^H\big\} \mathbb{E}\big\{ (\Delta\mathbf{H}_1^{\dagger} \hat{\mathbf{H}}_1(\Delta\mathbf{H}_1^{\dagger}\hat{\mathbf{H}}_1)^H) ^{-1}\big\}  \Big),
 \end{equation}

 \vspace{-0.2cm}\noindent
 where $(\text{a})$ is due to Lemma \ref{lemma4}, and $(\text{b})$ is due to Lemma \ref{lemma3} and the fact that the elements in $\bm{\mu}_k\bm{\mu}_k^H$ are independent to those in $\Delta\mathbf{H}_1^H(\hat{\mathbf{H}}_1^H)^{\dagger}\hat{\mathbf{H}}_1^{\dagger}\Delta\mathbf{H}_1$. Furthermore,
since the elements in $\mathbf{H}_1$ and $\Delta\mathbf{H}_1$ are independent, it follows that
 $ \mathbb{E}\big\{ \Delta\mathbf{H}_1^{\dagger} \hat{\mathbf{H}}_1(\Delta\mathbf{H}_1^{\dagger}\hat{\mathbf{H}}_1)^H\big\} = \mathbb{E} \big\{\Delta\mathbf{H}_1^{\dagger}\mathbb{E}\{{\mathbf{H}}_1{\mathbf{H}}_1^H\}(\Delta\mathbf{H}_1^{\dagger})^H\big\} + \mathbf{I}_N
\overset{\text{(a)}}{=}  Nl^{\text{UB}}_1l^{\text{IB}}\mathbb{E} \big\{(\Delta\mathbf{H}_1^H\Delta\mathbf{H}_1)^{-1}\big\} + \mathbf{I}_N \overset{\text{(b)}}{\approx}  \frac{Nl^{\text{UI}}_1l^{\text{IB}}p(N+1)+M\sigma^2} {M\sigma^2}\mathbf{I}_N$,
 where $(\text{a})$ is because $\mathbb{E}\{\mathbf{H}_1\mathbf{H}_1^H\} = Nl^{\text{UI}}_1l^{\text{IB}}\mathbf{I}_M$ which is obtained based on the reflected channel model in \eqref{A-11}, and (b) is due to the fact that  $\mathbb{E}\{\Delta\mathbf{H}_1^H\Delta\mathbf{H}_1\} = \frac{M\sigma^2}{p(N+1)}\mathbf{I}_N$ is a diagonal matrix. Since $ \mathbb{E}\big\{ \Delta\mathbf{H}_1^{\dagger} \hat{\mathbf{H}}_1(\Delta\mathbf{H}_1^{\dagger}\hat{\mathbf{H}}_1)^H\big\}$ is a diagonal matrix, we obtain  $\mathbb{E}\big\{ \big( \Delta\mathbf{H}_1^{\dagger} \hat{\mathbf{H}}_1(\Delta\mathbf{H}_1^{\dagger}\hat{\mathbf{H}}_1)^H\big)^{-1} \big\} \approx \big(\mathbb{E}\big\{ \Delta\mathbf{H}_1^{\dagger} \hat{\mathbf{H}}_1(\Delta\mathbf{H}_1^{\dagger}\hat{\mathbf{H}}_1)^H\big\} \big)^{-1}$
 according to Proposition \ref{PRO1}.
 As a result,  $\mathbf{P}_{\bm{\mu}_k,2}^{\text{a}}$ can be approximated by
 $\mathbf{P}_{\bm{\mu}_k,2}^{\text{a}}
 \approx  \frac{M\sigma^2\mathbb{E}\{\bm{\mu}_k^H \bm{\mu}_k\}}{Nl^{\text{UI}}_1l^{\text{IB}}p(N+1)+M\sigma^2}$. This  thus
completes the proof.

\vspace{0.1cm}
\append{\;\;Proof of Proposition \ref{Pro3}}\label{AppenC}

  First, by resorting to  Lemma \ref{lemma3}, $\mathbf{P}_{\bm{\mu}_k,1}^b$ can be transformed into  $\mathbf{P}_{\bm{\mu}_k,1}^b= \frac{\sigma^2}{p}\text{tr}\big(\mathbb{E}\big\{ (\hat{\mathbf{S}}^H_k\hat{\mathbf{S}}_k)^{-1}\big\}\big)$. Then, we define ${\mathbf{S}}_k \triangleq  [\text{diag}({{\boldsymbol{\theta}}}_{\xi_k+1}) {{\mathbf{H}}}_{1}^H,
\cdots,\text{diag}({{\boldsymbol{\theta}}}_{\xi_k+\gamma+1}){{\mathbf{H}}}_{1}^H]^H $ and it is readily seen that
the elements of $\mathbf{S}_k$ are independent with those of $\Delta \mathbf{S}_k$ due to the fact that $\mathbf{H}_1$ and $\Delta \mathbf{H}_1$ are uncorrelated.
 By replacing $\hat{\mathbf{S}}_k$ with ${\mathbf{S}}_k - \Delta{\mathbf{S}}_k$, we can obtain
$
\mathbb{E}\{\hat{\mathbf{S}}^H_k\hat{\mathbf{S}}_k\} = \mathbb{E}\{{{\mathbf{S}}}^H_k{{\mathbf{S}}}_k\} + \mathbb{E}\{\Delta{\mathbf{S}}^H_k\Delta{\mathbf{S}}_k\} = \frac{2pM(N+1)l^{\text{UI}}_1l^{\text{IB}}   + 2M\sigma^2}{p(N+1)}\mathbf{I}_N
$, which  is a diagonal matrix.
Besides, when $M$ is asymptotically large, we have  $\mathbb{E}\{(\hat{\mathbf{S}}^H_k\hat{\mathbf{S}}_k)^{-1}\}\approx ( \mathbb{E}\{\hat{\mathbf{S}}^H_k\hat{\mathbf{S}}_k\} )^{-1}$ according to Proposition \ref{PRO1}, then $\mathbf{P}_{\bm{\mu}_k,1}^b$  can be approximated by
$
\mathbf{P}_{\bm{\mu}_k,1}^b
\approx \frac{\sigma^2N(N+1)}{2pM(N+1)l^{\text{UI}}_1l^{\text{IB}}   + 2M\sigma^2}.
$
Second,  $\mathbf{P}_{\bm{\mu}_k,2}^b$ can be equivalently rewritten as  $\mathbf{P}_{\bm{\mu}_k,2}^b
= \text{tr}\big( \mathbb{E}\{\bm{\mu}_k\bm{\mu}_k^H\} \mathbb{E}\{ \tilde{\mathbf{S}}_k ^{-1}\}  \big)$ with $\tilde{\mathbf{S}}_k = \Delta\mathbf{S}_k^{\dagger} \hat{\mathbf{S}}_k(\Delta\mathbf{S}_k^{\dagger}\hat{\mathbf{S}}_k)^H$ similar to the derivation of  \eqref{Pmuk2a}.
Furthermore, since the elements of $\mathbf{S}_k$ and $\Delta \mathbf{S}_k$ are independent with each other, it follows that
$
\mathbb{E}\{\tilde{\mathbf{S}}_k\}
 = \mathbb{E} \big\{\big(\Delta\mathbf{S}^H_k (\mathbb{E}\{{{\mathbf{S}}}_k{{\mathbf{S}}}^H_k\})^{\dagger}\Delta\mathbf{S}_k\big)^{-1}\big\} + \mathbf{I}_N
 \overset{\text{(a)}}{\approx}  \big( \mathbb{E} \big\{\Delta\mathbf{S}_k^H (\mathbb{E}\{{{\mathbf{S}_k}}{{\mathbf{S}_k}}^H\})^{\dagger} \Delta\mathbf{S}_k\big\} \big)^{-1} + \mathbf{I}_N
$, where $(\text{a})$ is because $ \mathbb{E} \big\{\Delta\mathbf{S}_k^H (\mathbb{E}\{{{\mathbf{S}_k}}{{\mathbf{S}_k}}^H\})^{\dagger} \Delta\mathbf{S}_k\big\}$ is a diagonal matrix and Proposition \ref{PRO1} is then satisfied.  More specifically, according to Lemma \ref{lemma1}, the detailed expression of $ \mathbb{E}
\big\{\Delta\mathbf{S}_k^H (\mathbb{E}\{{{\mathbf{S}_k}}{{\mathbf{S}_k}}^H\})^{\dagger} \Delta\mathbf{S}_k\big\}$  can be obtained as
 \vspace{-0.4cm}\begin{equation}\label{E-8}
\begin{array}{c}
\mathbb{E} \big\{\Delta\mathbf{S}_k^H (\mathbb{E}\{{{\mathbf{S}_k}}{{\mathbf{S}_k}}^H\})^{\dagger} \Delta\mathbf{S}_k\big\}
= \frac{M\sigma^2}{p(N+1)l^{\text{UI}}_1l^{\text{IB}}}\left[
\begin{array}{cccc}
E_{Q,1}\mathbf{I}_M & \mathbf{0}_{M\times M} & \cdots & \mathbf{0}_{M\times M}\\
 \mathbf{0}_{M\times M} & E_{Q,2}\mathbf{I}_M &  \cdots & \mathbf{0}_{M\times M} \\
 \vdots & \vdots & \ddots & \vdots \\
  \mathbf{0}_{M\times M}  & \mathbf{0}_{M\times M}  & \cdots & E_{Q,\gamma}\mathbf{I}_\delta
\end{array}
\right],
\end{array}
 \end{equation}

 \vspace{-0.3cm}\noindent
where  $E_{Q,n} =  \frac{\gamma^2 + 2\gamma+2}{M(\gamma+1)^2}
-\frac{1}{\delta(\gamma+1)^2}$ if $n<\gamma$, and $E_{Q,n} =  -\frac{\gamma-1}{M(\gamma+1)^2}
+ \frac{\gamma^2 + 3\gamma }{\delta(\gamma+1)^2}$ otherwise. As a result, $\mathbb{E}\{\tilde{\mathbf{S}}_k\}$ (which is the sum of two diagonal matrices) is also a diagonal matrix, thus  $\mathbb{E}\big\{(\tilde{\mathbf{S}}_k)^{-1}\big\} \approx \big( \mathbb{E}\{\tilde{\mathbf{S}}_k\}\big)^{-1}$ holds based on Proposition
\ref{PRO1}. Accordingly,  $\mathbf{P}_{\bm{\mu}_k,2}^b$  can be approximated by
$
\mathbf{P}_{\bm{\mu}_k,2}^b  \approx \frac{(\delta \gamma^2 +2\gamma + 2\delta-M)\sigma^2\sum_{n=1}^{(\gamma-1)M}\mathbb{E}\{\mu_{k,n}\mu_{k,n}^*\}}{p\delta (\gamma +1)^2(N+1)l^{\text{UI}}_1l^{\text{IB}} + (\delta \gamma^2 +2\gamma + 2\delta-M)\sigma^2}
+ \frac{(M\gamma^2+(3M-1)\gamma +\delta)\sigma^2 \sum_{n=(\gamma-1)M+1}^N\mathbb{E}\{\mu_{k,n}\mu_{k,n}^*\}}{p\delta (\gamma +1)^2(N+1)l^{\text{UI}}_1l^{\text{IB}} +(M\gamma^2+(3M-1)\gamma +\delta)\sigma^2 }
$. This  thus completes the proof.

\vspace{0.0cm}
\append{\;\;Proof of Theorem \ref{PRO4}}\label{AppenD}

In the case of $M\ge N$,
 we first let $\epsilon_{d,\text{2P}}^{\text{a}} = \epsilon_{d,1,\text{2P}} + \sum_{k=2}^K \epsilon_{d,k,\text{2P}}^{\text{a}} $ denote the overall MSE for estimating all the direct channels under the proposed strategy, then we have
$
\epsilon_{d,\text{3P}} - \epsilon_{d,\text{2P}}^{\text{a}}
= \frac{M\sigma^2}{p}\big(\frac{1}{2}-\frac{1}{N+1}\big) \ge 0,
$
 which demonstrates  the 2PCE strategy is better than the 3PCE strategy
 when estimating the direct channels, and this performance gain holds regardless of the numbers of  BS antennas, reflecting elements and users.
 For the reflected channel associated with the typical user,  we have
$
\epsilon_{r,1,\text{3P}} - \epsilon_{r,1,\text{2P}} = \frac{(1+K)M\sigma^2}{pK} - \frac{NM\sigma^2}{p(N+1)}>0
$, i.e., the proposed channel estimation strategy can also achieve a better performance in this case.
Besides, focusing on the estimation of $\bm{\mu}_k$, it follows that
$
 \epsilon_{\bm{\mu}_k,\text{3P}}^{\text{a}}-\epsilon_{\bm{\mu}_k,\text{2P}}^{\text{a}}
=  A_{k,a} + B_{k,a} + C_{k,a} + D_{k,a}
$,
where $A_{k,a} = \frac{M\sigma^2\mathbb{E}\{\mu_{k,1}\mu_{k,1}^*\}}{pNl^{\text{BU}}_1l^{\text{BI}}\frac{NK}{K+N}+M\sigma^2}  - \frac{M\sigma^2\mathbb{E}\{\mu_{k,1}\mu_{k,1}^*\}}{pNl^{\text{BU}}_1l^{\text{BI}}(N+1)+M\sigma^2}
$,
$B_{k,a}=\frac{M\sigma^2\sum_{n=2}^N\mathbb{E}\{\mu_{k,n}\mu_{k,n}^*\}}{pN^2l^{\text{UI}}_1l^{\text{IB}}+M\sigma^2}  - \frac{M\sigma^2\sum_{n=2}^N\mathbb{E}\{\mu_{k,n}\mu_{k,n}^*\}}{pN(N+1)l^{\text{UI}}_1l^{\text{IB}}+M\sigma^2}
$, $ C_{k,a} = \frac{N\sigma^2}{pMl^{\text{UI}}_1l^{\text{IB}}\frac{NK}{(N-1)(K+1)}+M\sigma^2\frac{K}{(N-1)(K+1)}}
   - \frac{N\sigma^2}{2p(K-1)M l^{\text{UI}}_1l^{\text{IB}} + 2M\sigma^2\frac{K-1}{N+1}}$ and
   $D_{k,a} = \frac{(1+K)N\sigma^2}{pMNKl^{\text{UI}}_1l^{\text{IB}} +(K+N)M\sigma^2 }$.
   As can be observed,  $A_{k,a}$ is the difference of two fractions who have the same  numerator but different
denominators, since the denominator of the first fraction, i.e., $pNl^{\text{BU}}_1l^{\text{BI}}\frac{NK}{K+N}+M\sigma^2$, is smaller than that of the second fraction, i.e., $pNl^{\text{BU}}_1l^{\text{BI}}(N+1)+M\sigma^2$, we obtain
$A_{k,a}>0$.  Similarly, we have $B_{k,a}>0$ and $C_{k,a}>0$, together with the fact that $D_{k,a}$ is also larger than 0, we obtain
$\epsilon_{\bm{\mu}_k,\text{3P}}^{\text{a}} > \epsilon_{\bm{\mu}_k,\text{2P}}^{\text{a}}$.
To summarize, we can conclude that in the case of $M\ge N$,
the proposed 2PCE strategy
 outperforms the 3PCE strategy when M is asymptotically large.

\vspace{0.15cm}
\append{\;\;Proof of Theorem \ref{PRO5}}\label{AppenE}

 In the case of $M<N$, we first consider  the estimation of all the  direct channels.
 Let $\epsilon_{d,\text{2P}}^b = \epsilon_{d,1,\text{2P}} + \sum_{k=2}^K \epsilon_{d,k,\text{2P}}^b$ denote the overall MSE achieved by the proposed 2PCE strategy for estimating the direct channels, then
the corresponding MSE difference between the two considered  strategies is given by
$
\epsilon_{d,\text{3P}} - \epsilon_{d,\text{2P}}^b  
= \frac{M\sigma^2}{p}\big(\frac{N}{N+1} - \frac{K-1}{\gamma+1}
\big).
$
It can be observed that the sign of this MSE difference  depends on the  values of $K$ and $\gamma$, i.e.,  if $K < \gamma +2$ is satisfied, the performance of the 2PCE strategy  is better, and vice versa.
 As for the estimation of the scaling vectors,  the MSE difference  can be expressed as
$
\epsilon_{\bm{\mu}_k,\text{3P}}^b-\epsilon_{\bm{\mu}_k,\text{2P}}^b
= A_{k,b} + B_{k,b} + C_{k,b} + D_{k,b}+E_{k,b},
$
where $A_{k,b} = \frac{(1+K)N\sigma^2}{pKNM l^{\text{UI}}_1l^{\text{IB}}+(K+N)M\sigma^2}$, $ B_{k,b} = \frac{N\sigma^2}{pMl^{\text{UI}}_1l^{\text{IB}}\frac{NK}{(N-1)(K+1)}+M\sigma^2\frac{K}{(N-1)(K+1)}}
 - \frac{ N\sigma^2}{2pMl^{\text{UI}}_1l^{\text{IB}}   + \frac{2M\sigma^2}{N+1}}$, $C_{k,b} = \frac{\sigma^2 \mathbb{E}\{\mu_{k,1}\mu_{k,1}^*\}}{pl^{\text{UI}}_1l^{\text{IB}}\frac{KN}{K+N} +\sigma^2 } - \frac{\sigma^2 \mathbb{E}\{\mu_{k,1}\mu_{k,1}^*\}}{pl^{\text{UI}}_1l^{\text{IB}}\frac{(\delta \gamma^2 + 2\delta \gamma +\delta)(N+1)}{\delta \gamma^2 + 2\gamma + 2\delta -M} + \sigma^2}$, $D_{k,b}= \frac{\sigma^2\sum_{n=2}^{(\gamma-1)M} \mathbb{E}\{\mu_{k,n}\mu_{k,n}^*\}}{pNl^{\text{UI}}_1l^{\text{IB}} +\sigma^2}  -\frac{\sigma^2 \sum_{n=2}^{(\gamma-1)M} \mathbb{E}\{\mu_{k,n}\mu_{k,n}^*\}}{pl^{\text{UI}}_1l^{\text{IB}}\frac{(\delta \gamma^2 + 2\delta \gamma +\delta)(N+1)}{\delta \gamma^2 + 2\gamma + 2\delta -M} + \sigma^2}$ and
 $E_{k,b}=\frac{\sigma^2 \sum_{n=(\gamma-1)M+1}^N\mathbb{E}\{\mu_{k,n}\mu_{k,n}^*\}}{pl^{\text{UI}}_1l^{\text{IB}}
\frac{N}{M} + \sigma^2 } - \frac{\sigma^2 \sum_{n=(\gamma-1)M+1}^N\mathbb{E}\{\mu_{k,n}\mu_{k,n}^*\}}
{pl^{\text{UI}}_1l^{\text{IB}}
\frac{\delta \gamma^2 + 2\delta \gamma + \delta}{M\gamma^2 +(3M-1)\gamma + \delta} + \sigma^2}$.
 Notice that the numerators of the two fractions  in $B_{k,b}$ are the same while the denominator of the first fraction, i.e., $\frac{pNKMl^{\text{UI}}_1l^{\text{IB}}}{(N-1)(K+1)}+\frac{KM\sigma^2}{(N-1)(K+1)}$, is smaller than that of the second fraction, i.e., $2pMl^{\text{UI}}_1l^{\text{IB}}   + \frac{2M\sigma^2}{N+1}$, hence we have $ B_{k,b} >0$. Similarly, we can obtain  $C_{k,b} >0$, $D_{k,b} >0$ and $E_{k,b} >0$, therefore together with the fact that $A_{k,b}>0$,  we have  $\epsilon_{\bm{\mu}_k,\text{3P}}^b > \epsilon_{\bm{\mu}_k,\text{2P}}^b$.
 Besides, since $\epsilon_{r,1,\text{3P}} > \epsilon_{r,1,\text{2P}}$ holds according to Appendix D, we can infer that
 the proposed 2PCE strategy can outperform  the 3PCE strategy when estimating the reflected channels. The proof is thus completed.


\end{spacing}
\vspace{-0.3cm}

\begin{spacing}{1.43}
\bibliography{IRS_chaEst1}

\begin{thebibliography}{10}
\providecommand{\url}[1]{#1}
\csname url@samestyle\endcsname
\providecommand{\newblock}{\relax}
\providecommand{\bibinfo}[2]{#2}
\providecommand{\BIBentrySTDinterwordspacing}{\spaceskip=0pt\relax}
\providecommand{\BIBentryALTinterwordstretchfactor}{4}
\providecommand{\BIBentryALTinterwordspacing}{\spaceskip=\fontdimen2\font plus
\BIBentryALTinterwordstretchfactor\fontdimen3\font minus
  \fontdimen4\font\relax}
\providecommand{\BIBforeignlanguage}[2]{{%
\expandafter\ifx\csname l@#1\endcsname\relax
\typeout{** WARNING: IEEEtran.bst: No hyphenation pattern has been}%
\typeout{** loaded for the language `#1'. Using the pattern for}%
\typeout{** the default language instead.}%
\else
\language=\csname l@#1\endcsname
\fi
#2}}
\providecommand{\BIBdecl}{\relax}
\BIBdecl

\bibitem{Boccardi2014}
F.~Boccardi, R.~W. Heath, A.~Lozano, T.~L. Marzetta, and P.~Popovski, ``Five
  disruptive technology directions for {5G},'' \emph{IEEE Commun. Mag.},
  vol.~52, no.~2, pp. 74--80, Feb. 2014.

\bibitem{Shafi2017}
M.~Shafi, A.~F. Molisch, P.~J. Smith, T.~Haustein, P.~Zhu, P.~D. Silva,
  F.~Tufvesson, A.~Benjebbour, and G.~Wunder, ``{5G}: A tutorial overview of
  standards, trials, challenges, deployment, and practice,'' \emph{IEEE J. Sel.
  Areas Commun.}, vol.~35, no.~6, pp. 1201--1221, Jun. 2017.

\bibitem{Wu2020}
Q.~Wu and R.~Zhang, ``Towards smart and reconfigurable environment:
  {Intelligent} reflecting surface aided wireless network,'' \emph{IEEE Commun.
  Mag.}, vol.~58, no.~1, pp. 106--112, Jun. 2020.

\bibitem{Basar2019}
E.~Basar, M.~D. Renzo, J.~D. Rosny, M.~Debbah, M.~Alouini, and R.~Zhang,
  ``Wireless communications through reconfigurable intelligent surfaces,''
  \emph{IEEE Access}, vol.~7, pp. 116\,753--116\,773, Aug. 2019.

\bibitem{Qing2020}
Q.~Wu, S.~Zhang, B.~Zheng, C.~You, and R.~Zhang, ``{Intelligent} reflecting
  surface aided wireless communications: {A} tutorial,'' \emph{arXiv:
  2007.02759v2}, 2020.

\bibitem{Tan2018}
X.~Tan, Z.~Sun, D.~Koutsonikolas, and J.~M. Jornet, ``Enabling indoor mobile
  millimeter-wave networks based on smart reflect-arrays,'' in \emph{IEEE
  INFOCOM}, 2018, pp. 270--278.

\bibitem{MingMin2020}
M.~M. Zhao, Q.~Wu, M.~J. Zhao, and R.~Zhang, ``Exploiting amplitude control in
  intelligent reflecting surface aided wireless communication with imperfect
  {CSI},'' \emph{arXiv: 2005.07002v2}, 2020.

\bibitem{Larsson2020}
H.~Guo, Y.~Liang, J.~Chen, and E.~G. Larsson, ``Weighted sum-rate maximization
  for reconfigurable intelligent surface aided wireless networks,'' \emph{IEEE
  Trans. Wireless Commun.}, vol.~19, no.~5, pp. 3064--3076, Feb. 2020.

\bibitem{Aliu}
M.~M. Zhao, A.~Liu, Y.~Wan, and R.~Zhang, ``Two-timescale beamforming
  optimization for intelligent reflecting surface aided multiuser communication
  with {QoS} constraints,'' \emph{arXiv:2011.02237v1}, 2020.

\bibitem{RZhang2019}
Q.~Wu and R.~Zhang, ``Intelligent reflecting surface enhanced wireless network
  via joint active and passive beamforming,'' \emph{IEEE Trans. Wireless
  Commun.}, vol.~18, no.~11, pp. 5394--5409, Nov. 2019.

\bibitem{MMZhao2}
M.~M. Zhao, Q.~Wu, M.~J. Zhao, and R.~Zhang, ``Intelligent reflecting surface
  enhanced wireless network: {T}wo-timescale beamforming optimization,''
  \emph{IEEE Trans. Wireless Commun.}, DOI: 10.1109/TWC.2020.3022297, 2020.

\bibitem{Mishra2019}
D.~Mishra and H.~Johansson, ``Channel estimation and low-complexity beamforming
  design for passive intelligent surface assisted {MISO} wireless energy
  transfer,'' in \emph{IEEE ICASSP}, 2019, pp. 4659--4663.

\bibitem{9039554}
Y.~{Yang}, B.~{Zheng}, S.~{Zhang}, and R.~{Zhang}, ``Intelligent reflecting
  surface meets {OFDM}: {P}rotocol design and rate maximization,'' \emph{IEEE
  Trans. Commun.}, vol.~68, no.~7, pp. 4522--4535, Jul. 2020.

\bibitem{Jensen2019}
T.~L. Jensen and E.~D. Carvalho, ``An optimal channel estimation scheme for
  intelligent reflecting surfaces based on a minimum variance unbiased
  estimator,'' in \emph{IEEE ICASSP}, May 2020, pp. 5000--5004.

\bibitem{OFDM2020}
B.~Zheng and R.~Zhang, ``Intelligent reflecting surface-enhanced {OFDM}:
  {C}hannel estimation and reflection optimization,'' \emph{IEEE Wireless
  Commun. Lett.}, vol.~9, no.~4, pp. 512--522, Apr. 2020.

\bibitem{Tellambura2019}
J.~Lin, G.~Wang, R.~Fan, T.~A. Tsiftsis, and C.~Tellambura, ``Channel
  estimation for wireless communication systems assisted by large intelligent
  surfaces,'' \emph{arXiv:1911.02158v1}, 2019.

\bibitem{CYou2019}
C.~You, B.~Zheng, and R.~Zhang, ``{C}hannel estimation and passive beamforming
  for intelligent reflecting surface: {D}iscrete phase shift and progressive
  refinement,'' \emph{{IEEE} J. Select. Areas Commun.}, vol.~38, no.~11, pp.
  2604--2620, Nov. 2020.

\bibitem{He2019}
Z.~He and X.~Yuan, ``Cascaded channel estimation for large intelligent
  metasurface assisted massive {MIMO},'' \emph{IEEE Wireless Commun. Lett.},
  vol.~9, no.~2, pp. 210--214, Oct. 2020.

\bibitem{Jie2019}
J.~Chen, Y.~Liang, H.~V. Cheng, and W.~Yu, ``Channel estimation for
  reconfigurable intelligent surface aided multi-user {MIMO} systems,''
  \emph{arXiv:1912.03619v1}, 2019.

\bibitem{9149146}
Z.~{Wan}, Z.~{Gao}, and M.~{Alouini}, ``Broadband channel estimation for
  intelligent reflecting surface aided mmwave massive {MIMO} systems,'' in
  \emph{IEEE ICC}, Jun. 2020, pp. 1--6.

\bibitem{9214532}
W.~{Zhang}, J.~{Xu}, W.~{Xu}, D.~W.~K. {Ng}, and H.~{Sun}, ``Cascaded channel
  estimation for {IRS}-assisted mmwave multi-antenna with quantized
  beamforming,'' \emph{IEEE Commun. Lett.}, DOI: 10.1109/LCOMM.2020.3028878,
  2020.

\bibitem{arXiv1912}
Z.~Wang, L.~Liu, and S.~Cui, ``Channel estimation for intelligent reflecting
  surface assisted multiuser communications: {F}ramework, algorithms, and
  analysis,'' \emph{IEEE Trans. Wireless Commun.}, vol.~19, no.~10, pp.
  6607--6620, Jun. 2020.

\bibitem{McKay2005}
M.~R. McKay and I.~B. Collings, ``General capacity bounds for spatially
  correlated {R}ician {MIMO} channels,'' \emph{IEEE Trans. Inf. Theory},
  vol.~51, no.~9, pp. 3121--3145, Sept. 2005.

\bibitem{Loyka2001}
S.~L. Loyka, ``Channel capacity of {MIMO} architecture using the exponential
  correlation matrix,'' \emph{IEEE Commun. Lett.}, vol.~5, no.~9, pp. 369--371,
  Sept. 2001.

\bibitem{YiWei2019}
Y.~Wei, M.~M. Zhao, M.~J. Zhao, and M.~Lei, ``Learned conjugate gradient
  descent network for massive {MIMO} detection,'' \emph{IEEE Trans. Signal
  Process.}, vol.~68, pp. 6336--6349, Nov. 2020.

\bibitem{Choi2014}
J.~Choi and D.~J. Love, ``Bounds on eigenvalues of a spatial correlation
  matrix,'' \emph{IEEE Commun. Lett.}, vol.~18, no.~8, pp. 1391--1394, Aug.
  2014.

\bibitem{arXiv01301}
Q.~Nadeem, H.~Alwazani, A.~Kammoun, A.~Chaaban, M.~Debbah, and M.~Alouini,
  ``Intelligent reflecting surface assisted multi-user {MISO} communication:
  {C}hannel estimation and beamforming design,'' \emph{IEEE OJ-COMS}, vol.~1,
  pp. 661--680, May 2020.

\bibitem{NA1993}
R.~L. Burden and J.~D. Faires, \emph{Numerical Analysis (5th ed.)}.\hskip 1em
  plus 0.5em minus 0.4em\relax Boston: Prindle, 1993.

\bibitem{Livan2018}
G.~Livan, M.~Novaes, and P.~Vivo, \emph{Introduction to Random Matrices Theory
  and Practice}.\hskip 1em plus 0.5em minus 0.4em\relax SpringerBriefs in
  Mathematical Physics, 2018.

\bibitem{GIORGIO}
G.~Cipolloni and L.~Erdos, ``Fluctuations for linear eigenvalue statistics of
  sample covariance random matrices,'' \emph{arXiv:1806.08751v2}, 2018.

\bibitem{COVARIANCE2009}
T.~Cai, C.~Zhang, and H.~Zhou, ``Optimal rates of convergence for covariance
  matrix estimation,'' \emph{The Annals of Statistics}, vol.~38, no.~4, pp.
  2118--2144, 2010.

\bibitem{Bernstein2006}
D.~Bernstein, \emph{Matrix Mathematics}.\hskip 1em plus 0.5em minus 0.4em\relax
  UK: Princeton University Press, 2005.

\bibitem{Hartwing1976}
R.~E. Hartwing, ``Block generalized inverses,'' \emph{Arch. Rational Mech.
  Anal.}, no.~61, pp. 197--251, 1976.

\bibitem{Meyer2000}
P.~Babington, \emph{Matrix Analysis and Applied Linear Algebra}.\hskip 1em plus
  0.5em minus 0.4em\relax USA: Society for Industrial and Applied Mathematics,
  2000.

\end{thebibliography}
\bibliographystyle{IEEETran}
\end{spacing}

\end{document}